\documentstyle[pra,aps]{revtex}
\begin{document}
\draft
\title{Gauge Unified Theories without the 
Cosmological Constant Problem}
\author{E. I. Guendelman\thanks{GUENDEL@BGUmail.BGU.AC.IL} and
        A. B. Kaganovich \thanks{ALEXK@BGUmail.BGU.AC.IL}}
\address{Physics Department, Ben Gurion University of the Negev,
   Beer  Sheva 84105, Israel}
\maketitle
\begin{abstract}
We study gauge theories in the context of a gravitational theory without the 
cosmological constant problem (CCP). The theory is based on the requirement 
that the measure of integration in the action is not necessarily $\sqrt{-g}$ 
but it is determined dynamically through additional degrees of freedom. 
Realization of these ideas in the framework of the first order formalism solves
the CCP. Incorporation of a condensate of a four index field strength 
allows, after a conformal transformation to the Einstein frame, to represent 
the system  of gravity and matter in the standard GR form. Now, however, 
the  {\em effective} potential vanishes at a vacuum state due to the exact 
balance to zero of the gauge fields condensate and the original scalar fields 
potential. As a result it is possible to combine the solution of the CCP with: 
a) inflation and transition to a $\Lambda =0$ phase without fine tuning after a 
reheating period; b) spontaneously broken gauge unified theories (including 
fermions). The model opens new possibilities for a solution of the hierarchy 
problem.  

\vspace{1cm}

\vspace{1cm}

\end{abstract}
 \pagebreak
\section{Introduction}

\bigskip

The cosmological constant problem in the context of general relativity 
(GR) can be explained as follows. In GR one can introduce such a constant 
or one may set it to zero. The problem is that after further 
investigation of elementary particle theory, we discover new phenomena 
like radiative corrections, the existence of condensates, etc. each of 
which contributes to the vacuum energy. In order to have a resulting zero 
or extremely small cosmological constant as required by observations of 
the present day universe , one would have to carefully fine tune 
parameters in the Lagrangian so that all of these contributions more or 
less exactly cancel. This question has captured the attention of many 
authors because, among other things, it could be a serious indication 
that something fundamental has been missed in our standard way of 
thinking about field theory and the way it must couple to gravity. For a 
review of this problem see \cite{CC}.

The situation is made even more serious if one believe in the existence 
of an inflationary phase for the early universe, where the vacuum energy 
plays an essential role. The question is then: what is so special about 
the present vacuum state which was not present in the early universe? In 
this paper we are going to give an answer this question. 

As it is well known, in nongravitational physics the origin from which we
measure energy is not important. For example in nonrelativistic mechanics
a shift in the potential $V\rightarrow V+constant$ does not lead to any
consequence in the equations of motion. In the GR the situation changes 
dramatically. There {\em all} the
energy density, including the origin from which we measure, affects the
gravitational dynamics.

This is quite apparent when GR is formulated from a variational approach.
There the action is
\begin{equation}
S=\int\sqrt{-g}Ld^{4}x
\label{I1}
\end{equation}
\begin{equation}
L=-\frac{1}{\kappa}R(g)+L_{m}
\label{I2}
\end{equation}
where $\kappa=16\pi G$, $R(g)$ is the Riemannian scalar curvature of the
4-dimensional space-time with metric $g_{\mu\nu}$, $g\equiv
Det(g_{\mu\nu})$ and $L_{m}$ is the matter Lagrangian density.
It is apparent now that the shift of the Lagrangian density $L$,
\quad $L\rightarrow L+C$,\quad $C=const$ is not a symmetry of the action 
(\ref{I1}).
Instead, it leads to an additional piece in the action of the form
$C\int\sqrt{-g}d^{4}x$ which contributes to the equations of motion and
in particular generates a so called "cosmological constant term" in the
equations of the gravitational field. 

In Refs. \cite{GK1}-\cite{GK3} we have developed an approach where the
cosmological constant problem is treated as the absence of gravitational
effects of a possible constant part of the Lagrangian density. The basic
idea is that the measure of integration in the action principle is not
necessarily $\sqrt{-g}$ but it is allowed to "float" and to be determined
dynamically through additional degrees of freedom. In other words the
floating measure is not from first principles related to $g_{\mu\nu}$, 
although relevant equations will in general allow to solve for the new 
measure in terms of other fields of the theory ($g_{\mu\nu}$ and matter 
fields). This theory is based on the demand that such measure respects  
{\em the principle of
non gravitating vacuum energy} (NGVE principle) which states that the
Lagrangian density $L$ can be changed to $L+constant$ without affecting
the dynamics. This requirement is imposed in order to offer a new
approach for the solution of the cosmological constant problem.
Concerning the theories based on the NGVE principle we will refer to
them  as NGVE-theories. 

The invariance $L\longrightarrow L+constant$ for the action is
achieved if the measure of integration in the action is a total
derivative, so that
to an infinitesimal hypercube in 4-dimensional space-time $x_{0}^{\mu}\leq
x^{\mu}\leq x_{0}^{\mu}+dx^{\mu}$, $\mu =0,1,2,3$ we associate a volume
element $dV$ which is: (i) a total derivative, (ii) it is proportional to
$d^{4}x$ and (iii) $dV$ is a general coordinate invariant. The usual choice,
$\sqrt{-g} d^{4}x$ does not satisfy condition (i).

The conditions (i)-(iii) are satisfied \cite{GK1}, \cite{GK2} if
the measure corresponds to the integration in the space of the four 
scalar fields  $\varphi_{a}, (a=1,2,3,4)$, that is
\begin{equation}
dV =
d\varphi_{1}\wedge
d\varphi_{2}\wedge
d\varphi_{3}\wedge
d\varphi_{4}\equiv\frac{\Phi}{4!}d^{4}x
\label{dV}
\end{equation}
where
\begin{equation}
\Phi \equiv \varepsilon_{a_{1}a_{2}a_{3}a_{4}}
\varepsilon^{\mu\nu\lambda\sigma}
(\partial_{\mu}\varphi_{a_{1}})
(\partial_{\nu}\varphi_{a_{2}})
(\partial_{\lambda}\varphi_{a_{3}})
(\partial_{\sigma}\varphi_{a_{4}}).
\label{Fi}
\end{equation}
Notice that the measure which was discussed in
Introduction, is a particular realization of the NGVE-principle (for
other possible realization which leads actually to the same results, see
Sec.I of Ref \cite{GK3}).
For deeper discussion of the geometrical meaning of this realization of the 
measure see Ref.\cite{Hehl} 

The total action  is defined as follows
\begin{equation}
S =\int L\Phi d^{4}x
        \label{Action}
\end{equation}
where $L$ is a total Lagrangian density. We assume in what follows that 
$L$ does not contain explicitly the measure fields, that is the fields
$\varphi_{a}$ by means of which $\Phi$ is defined. If this is satisfied, 
an infinite dimensional symmetry appears (see Sec.II).

Introducing independent degrees of freedom related to the measure we 
arrive naturally at a conception that all possible degrees of freedom that 
can appear should be considered as such. This is why we expect that the 
first order formalism, where the affine connection is {\em not} assumed to 
be the Christoffel coefficients in general should be preferable to the 
second order formalism where this assumption is made. 

In fact, it is found that the NGVE theory in the context of the first order 
formalism does indeed provide a solution of the cosmological constant 
problem \cite{GK2}, while this is not the case when using the second 
order formalism. 

The simplest example \cite{GK2},\cite{GK3} (see also Sec. IIIB) where these 
ideas can be tested is that of a matter Lagrangian described by a single 
scalar field with a nontrivial potential. In this case the variational 
principle leads to a constraint which implies the vanishing of the 
effective vacuum energy in any possible allowed configuration of the 
scalar field. These allowed configurations are however constant values at 
the extrema of the scalar field potential and an integration constant 
that results from the equations of motion has the effect of exactly 
canceling the value of the potential at these points. So, the scalar 
field is forced to be a constant and hence  
the theory has no nontrivial dynamics for the scalar field.

In this case the measure (\ref{Fi}) is not determined by the equations of 
motion. In fact a local symmetry (called "Local Einstein Symmetry" (LES)) 
exists which allows us to choose the measure $\Phi$ to be o
whatever we 
want. In 
particular $\Phi=\sqrt{-g}$ can be chosen and in this case the theory 
coincides in the vacuum with GR with $\Lambda =0$.

A richer structure is obtained if a four index field strength which 
derives from a three index potential is allowed in the theory. The 
introduction of this term breaks the LES mentioned above. In this case, 
the constraint that the theory provides, allows  to solve for the measure 
in terms of $\sqrt{-g}$ {\em and} the matter fields of the theory. The 
equations can be written in a form that resemble those of the Einstein 
theory by the use of a conformal transformation (or in equivalent language 
by going to the effective Einstein frame).

Then the theory which contains a scalar field shows a remarkable feature: 
the effective potential of the scalar field that one obtains in the 
Einstein conformal frame 
is that which generally allows for an inflationary phase which evolves at 
a later stage, 
without fine tuning, to a  vacuum of the theory  with zero cosmological 
constant \cite{GK4}. This will be discussed in Sec.IVA.

The 4-index field strength also allows  for a Maxwell-type dynamics of 
gauge fields and of massive fermions. In this case the effective 
coupling constants of gauge theories and the fermion masses depend on a 
dimensionless integration constant $\omega$
associated with the integration of the equation of motion for the 
4-index field strength. This integration constant $\omega$, together 
with another integration constant $M$ (that goes together with the 
original scalar field potential $V(\varphi)$, to form the combination 
$V+M$) determines the scalar field potential which has an 
absolute minimum  with zero effective vacuum energy (fine 
tuning is not necessary). Furthermore, in the context of cosmology, where 
the scalar field plays the role of inflaton,  $\omega$ and $M$ also 
govern the parameters of the inflationary picture present in this model.

In the most simple version of the theory with a scalar field and 
4-index field strength, the scalar field has nontrivial cosmological 
dynamics, however at the absolute minimum $V+M=0$ (if no fine tuning is 
made), the 
mass of the scalar is generically infinite. Additional problems in this 
simplest model are related to an appearance of  nonrenormalizable 
couplings between gauge and scalar fields.  All these difficulties can be 
overcomed (without changing the advantages described above) by 
a certain sort of unification of all gauge fields, together with 
modification of  the kinetic term of the gauge sector in the original  
Lagrangian density. It is argued that some of these modifications are 
natural if we believe in the existence of gauge fields condensates in 
the vacuum. In 
Sec. V, we implement this in the framework of two different models: 
"model with a critical limit" and "model with persistent 
condensate". These models allow to include fermions and to provide 
mass generation for them (see Sec. VI). In the model with persistent 
condensate in the $\Lambda =0$ vacuum there is an exact balance between 
the integration constant $M$, the original potential and the 
contribution from the gauge condensate.  An explicit construction of
unified gauge theory ($SU(2)\times U(1)$ as an example) based on 
these ideas and keeping all the above mentioned advantages is 
presented in Sec.VII.

\section{General features of the NGVE theory in the first order formalism}

\bigskip

We assume that the total Lagrangian density $L$ in Eq. (\ref{Action}) does 
not contain the measure fields $\varphi_{a}$, that 
is the fields by means of which the measure $\Phi$ is defined. If this 
condition is
satisfied then the theory has an additional symmetry. In fact, the action 
(\ref{Action}) is invariant under the infinitesimal shift of the
fields $\varphi_{a}$ by an arbitrary infinitesimal function of the total
Lagrangian density $L$, that is \cite{GK1}
\begin{equation}
\varphi^{\prime}_{a}=\varphi_{a}+\epsilon g_{a}(L),\  \epsilon\ll 1
        \label{LP}
\end{equation}

Our choice for the total Lagrangian
density is
\begin{equation}
L=-\frac{1}{\kappa}R(\Gamma,G)+L_{m}
\label{L1}
\end{equation}
where $L_{m}$ is the matter Lagrangian density and $R(\Gamma,G)$ is the
scalar  curvature which in the first order formalism in the framework of 
the Metric-Affine theory \cite{HN} is defined as follows
 \begin{equation}
R(\Gamma,g)=g^{\mu\nu}R_{\mu\nu}(\Gamma)
\label{R}
\end{equation}

\begin{equation}
R_{\mu\nu}(\Gamma)=R^{\lambda}_{\mu\nu\lambda}(\Gamma)
\label{RAB}
\end{equation}

\begin{equation}
R^{\lambda}_{\mu\nu\sigma}(\Gamma)\equiv 
\Gamma^{\lambda}_{\mu\nu ,\sigma}-\Gamma^{\lambda}_{\mu\sigma ,\nu}+
\Gamma^{\lambda}_{\alpha\sigma}\Gamma^{\alpha}_{\mu\nu}-
\Gamma^{\lambda}_{\alpha\nu}\Gamma^{\alpha}_{\mu\sigma}
\label{RABCD}
\end{equation}  
where $\Gamma^{\lambda}_{\mu\nu}$ are the connection coefficients which 
have to be obtained from the variational principle.

Equations that originate from the variation of the action (\ref{Action}) 
with respect to the measure fields $\varphi_{a}$, are  
\begin{equation}
A^{\mu}_{a}\partial_{\mu}\lbrack -\frac{1}{\kappa}R(\Gamma,g)+
L_{m}\rbrack =0
\label{FEM}
\end{equation}
where $A^{\mu}_{a}=\varepsilon_{bcda}
\varepsilon^{\nu\lambda\sigma\mu}
(\partial_{\nu}\varphi_{b})
(\partial_{\lambda}\varphi_{c})
(\partial_{\sigma}\varphi_{d})$. Since
$A_{a}^{\mu}\partial_{\mu}\varphi_{a^{\prime}}=4^{-1}\delta_{aa^{\prime}}\Phi$
it follows that  $Det (A^{\mu}_{a}) =
\frac{4^{-4}}{4!}\Phi^{3}$, so that if $\Phi\neq 0$, it follows from 
Eq.(\ref{FEM})
\begin{equation}
 -\frac{1}{\kappa}R(\Gamma,g)+
L_{m}=M=constant
\label{II1}
\end{equation}

Let us now study equations that originate from variation with respect to 
$g^{\mu\nu}$. For simplicity we present here the calculations for the 
case where there are no fermions. Performing the variation with respect 
to $g^{\mu\nu}$ we get
\begin{equation}
 -\frac{1}{\kappa}R_{\mu\nu}(\Gamma)+
\frac{\partial L_{m}}{\partial g^{\mu\nu}}=0
\label{II2}
\end{equation}

Contracting Eq.(\ref{II2}) with $g^{\mu\nu}$ and making use 
Eq.(\ref{II1}) we get
\begin{equation}
g^{\mu\nu}\frac{\partial (L_{m}-M)}{\partial g^{\mu\nu}}-
(L_{m}-M)
=0
\label{II3}
\end{equation}

This equation is a constraint since generically $L_{m}$ contains only the 
fields and their first derivatives. A similar constraint is achieved by 
using the vierbein - spin-connection (VSC) formalism (see Appendix B). 
Notice that if $L_{m}-M$ is homogeneous of degree one in $g^{\mu\nu}$ it 
satisfies the constraint (\ref{II3}) automatically (that is without using 
equations of motion). 

\bigskip
\section{The theory for the vacuum, for some simple models 
\protect\\
 and Local Einstein Symmetry}

\bigskip
\subsection{The vacuum case}
In the vacuum, when we choose $L_{m}=0$ in Eq.(\ref{L1}), it follows from 
Eq.(\ref{II2})
\begin{equation}
R^{\mu\nu}(\Gamma)
=0
\label{III1}
\end{equation}

Eq.(\ref{II1}) implies then that the integration constant $M=0$. Adding 
an arbitrary constant $C$ to $L_{m}$ does not change the resulting 
Eq.(\ref{III1}) since, as we see from Eqs. (\ref{II1}) and (\ref{II2}) or 
from the constraint (\ref{II3}), the integration constant $M$ has to 
compensate the constant $C$.

To clarify the sense of Eq.(\ref{III1}) we need the connection 
coefficients $\Gamma^{\lambda}_{\mu\nu}$. Varying the action (\ref{Action})
with $L=-\frac{1}{\kappa}R(\Gamma ,g)$ with respect to 
$\Gamma^{\lambda}_{\mu\nu}$,  we get
\begin{eqnarray}
-\Gamma^{\lambda}_{\mu\nu}-
\Gamma^{\alpha}_{\beta\mu}g^{\beta\lambda}g_{\alpha\nu}+
\delta^{\lambda}_{\nu}\Gamma^{\alpha}_{\mu\alpha}
+\delta^{\lambda}_{\mu}g^{\alpha\beta}\Gamma^{\gamma}_{\alpha\beta}
g_{\gamma\nu}-
\nonumber\\
g_{\alpha\nu}\partial_{\mu}g^{\alpha\lambda}+
\delta^{\lambda}_{\mu}g_{\alpha\nu}\partial_{\beta}g^{\alpha\beta}-
\delta^{\lambda}_{\nu}\frac{\Phi,_{\mu}}{\Phi}+
\delta^{\lambda}_{\mu}\frac{\Phi,_{\nu}}{\Phi}
=0.
\label{GAM1}
\end{eqnarray}

We will look for the solution 
of the form
\begin{equation}
\Gamma^{\lambda}_{\mu\nu}=\{ ^{\lambda}_{\mu\nu}\}+\Sigma^{\lambda}_{\mu\nu}
\label{GAM2}
\end{equation}
where $\{ ^{\lambda}_{\mu\nu}\}$  are the Christoffel's connection 
coefficients. Then $\Sigma^{\lambda}_{\mu\nu}$ satisfies the equation
\begin{equation}
-\sigma,_{\lambda}g_{\mu\nu}+\sigma,_{\mu}g_{\nu\lambda}-
g_{\nu\alpha}\Sigma^{\alpha}_{\lambda\mu}-
g_{\mu\alpha}\Sigma^{\alpha}_{\nu\lambda}+
g_{\mu\nu}\Sigma^{\alpha}_{\lambda\alpha}+
g_{\nu\lambda}g_{\alpha\mu}g^{\beta\gamma}\Sigma^{\alpha}_{\beta\gamma}=0
\label{S1}
\end{equation}
where
\begin{equation}
\sigma\equiv\ln\chi, \hspace{1.5cm} \chi\equiv\frac{\Phi}{\sqrt{-g}}
\label{ski}
\end{equation}

The general solution of eq.
(\ref{S1}) is
\begin{equation}
\Sigma^{\alpha}_{\mu\nu}=\delta^{\alpha}_{\mu}\lambda,_{\nu}+
\frac{1}{2}(\sigma,_{\mu}\delta^{\alpha}_{\nu}-
\sigma,_{\beta}g_{\mu\nu}g^{\alpha\beta})
\label{S2}
\end{equation}
where $\lambda$ is an arbitrary function which appears due to the 
existence of the Einstein - Kaufman $\lambda$-symmetry (see \cite{EK}, 
\cite{GK3} and Appendix A): the curvature tensor (\ref{RABCD}) is 
invariant under the $\lambda$- transformation
\begin{equation}
\Gamma^{\prime
\alpha}_{\mu\nu}(\lambda, 
\sigma)=\Gamma^{\alpha}_{\mu\nu}+\delta^{\alpha}_{\mu}\lambda,_{\nu} 
\label{Gamal} 
\end{equation}
Although this symmetry was discussed in Ref.\cite{EK} in a very specific 
unified theory, it turns out that $\lambda$- symmetry has a wider range 
of validity and in particular it is useful in our case. 

If we choose the gauge $\lambda=\frac{\sigma}{2}$, then the antisymmetric 
part of $\Sigma^{\alpha}_{\mu\nu}$ disappears and we get finally
\begin{equation}
\Sigma^{\alpha}_{\mu\nu}(\sigma)=\frac{1}{2}(\delta^{\alpha}_{\mu}\sigma,_{\nu}
+\delta^{\alpha}_{\nu}\sigma,_{\mu}-
\sigma,_{\beta}g_{\mu\nu}g^{\alpha\beta})
\label{S3}
\end{equation}

In the vacuum, the $\sigma$-contribution (\ref{S3}) to the
nonmetricity (see Appendix A) can be eliminated. This is because in the
vacuum the action (\ref{Action}), (\ref{L1}) is invariant under the local 
Einstein symmetry (LES)
 \begin{equation}
g_{\mu\nu}(x)=J^{-1}(x)g^{\prime}_{\mu\nu}(x)
\label{ES1}
\end{equation}
\begin{equation}
\Phi(x)=J^{-1}(x)\Phi^{\prime}(x)
\label{ES2}
\end{equation}

The transformation (\ref{ES2})
can be the result of a diffeomorphism
$\varphi_{a}\longrightarrow\varphi^{\prime}_{a}=
\varphi^{\prime}_{a}(\varphi_{b})$ in the space of the scalar fields
$\varphi_{a}$ (see Ref. \cite{GK2}). Then $J=
Det(\frac{\partial\varphi^{\prime}_{a}}{\partial\varphi_{b}})$.

Notice that even when we are not in the vacuum, but the matter Lagrangian 
density $L_{m}$ satisfies the constraint (\ref{II3}) automatically (that 
is $L_{m}$ is homogeneous of degree one in $g^{\mu\nu}$, up to irrelevant 
additive constant) then the total action  (\ref{Action}), (\ref{L1}) 
possesses LES too. For examples see \cite{GK2},\cite{GK3}.  

 For $J=\chi$ we get $\chi^{\prime}\equiv 1$,
$\Sigma^{\prime \alpha}_{\mu\nu}(\sigma)\equiv 0$ and $\Gamma^{\prime 
\alpha}_{\mu\nu}= \{ ^{\alpha}_{\mu\nu}\}^{\prime}$, where
$\{ ^{\alpha}_{\mu\nu}\}^{\prime}$
 are the Christoffel's coefficients corresponding
to the new metric $g^{\prime}_{\mu\nu}$.
In terms of the new metric $g^{\prime}_{\mu\nu}$, the curvature 
(\ref{RABCD}) becomes the Riemannian curvature and therefore Eq. (\ref{III1})
is equivalent to the vacuum Einstein's equation with zero cosmological 
constant.

\bigskip
\subsection{Single scalar field with a nontrivial potential}

Now let us consider the cases when the constraint (\ref{II3}) is not satisfied
without restrictions on the dynamics of the matter fields. Nevertheless,
the constraint (\ref{II3}) holds as a consequence of the variational
principle in any situation.

A simple case where the constraint (\ref{II3}) is not automatic is the case
of a single scalar field with a nontrivial potential $V(\varphi)$
\begin{equation}
L_{m}= \frac{1}{2}\varphi_{,\alpha}\varphi^{,\alpha}-V(\varphi)
 \label{III2}
 \end{equation}
In this model, the kinetic part of the action possesses LES and satisfies 
the constraint automatically since 
$\frac{1}{2}\varphi_{,\alpha}\varphi^{,\alpha}$ is homogeneous of degree 
one in 
$g^{\mu\nu}$. The potential part apparently does not satisfy the LES and 
as a result of this the constraint (\ref{II3}) implies
\begin{equation}
V(\varphi)+M=0
 \label{CS}
 \end{equation}
Therefore we conclude that, provided $\Phi\neq 0$, there is no dynamics for
the theory of a single scalar field, since constraint (\ref{CS}) forces
this scalar field to be a constant. 

 The constraint (\ref{CS}) has to be
solved together with the equation of motion
\begin{equation}
\Box\varphi+\sigma ,_{\mu}\varphi ^{,\mu}+\frac{\partial 
V}{\partial\varphi}=0,
 \label{SE}
 \end{equation}
where $\sigma =\ln\chi$. From eqs.(\ref{CS}) and (\ref{SE}) we conclude
that the $\varphi$ -field has to be
located at an extremum of the potential
$V(\varphi)$. Since the constraint(\ref{CS}) eliminates the dynamics of the
scalar field $\varphi$, we cannot really say that we have a situation where
the LES (\ref{ES1}),(\ref{ES2}) is actually broken, since after solving
the constraint together with the equation of motion (i.e. on the mass 
shell) the symmetry remains true.

Taking into account that $\varphi =constant$ and 
$L_{m}-M=-(V(\varphi)+M)=0$, we 
see from Eqs. (\ref{II1})-(\ref{II3}) that $R_{\mu\nu}(\Gamma,g)=0$. As 
we have seen in Sec. IIIA, the $\sigma$ contribution to the 
connection can be eliminated in the vacuum by the transformations 
(\ref{ES1}), (\ref{ES2}). Notice that since $\phi =constant$, the 
single scalar field $\phi$ part of the Lagrangian density acts as 
an arbitrary constant. As we have seen also in Sec. IIIA , this 
situation is undistinguishable from the vacuum case. Then repeating the 
LES transformation of the end of Sec. IIIA, we see that in 
terms of the new metric $g^{\prime}_{\mu\nu}$, 
the tensor $R_{\mu\nu}(\Gamma,g)$ becomes the usual Ricci tensor 
$R_{\mu\nu}(g^{\prime})$ of the Riemannian space-time with the metric 
$g^{\prime}_{\mu\nu}$.  
Therefore we conclude that {\em for the case of a single scalar field with a 
nontrivial potential, the theory is equivalent to the Einstein's GR with 
the zero cosmological constant}. As a consequence, in this simple model, 
among maximally symmetric solutions, only Minkowski space is a solution. 
The absence of deSitter space as a solution makes us suspect that the 
NGVE theory is inconsistent with the idea of inflation. This however is 
not true as we will see in Sec.IV.

\bigskip
\section{Four index field strength condensate as an universal governor. 
\protect\\
Simple models.}
\bigskip

As we have mentioned, one of the biggest puzzles of modern physics is
what is referred to as the
"cosmological constant problem", i.e. the absence of a possible constant
part of the vacuum energy in the present day universe \cite{CC}. On the
other hand, many questions in modern
cosmology appear to be solved by the so called "inflationary model" which
makes use of a big effective cosmological constant in the
early universe \cite{Gut}. A possible conflict between a successful
resolution of the cosmological constant problem and the existence of
an inflationary phase could be a "potential Achilles heel for the
scenario" as has been pointed out\cite{KT}. Here we will show (see also 
\cite{GK4}) that indeed there is no conflict between the existence of an
inflationary phase and the disappearance of the cosmological constant  in
the later phases of cosmological evolution (without the need of fine
tuning). The four index field strength condensate plays a crucial role 
for this.

Another problem related to the NGVE theory consists of the very strong 
restriction which constraint (\ref{II3}) dictates on the matter models 
which generally do not satisfy the LES. This makes the incorporation of 
 fermion masses and gauge fields not straightforward (see Refs. 
\cite{GK2}, \cite{GK3}). In what follows we will see, however, that the 
incorporation of the four index-field strength in four dimensional 
space-time turns the constraint into an equation for $\chi$- field. After 
solving this constraint we obtain well defined matter models. However, 
as we will see, undesirable problems appear nevertheless even in this 
model: (a) the mass of the scalar field turns out to be infinite; 
(b) nonminimal nonrenormalizable couplings appear at very high energies. 
In Sec.V we will see that treating all gauge field strengths, including 
the four-index one in a unified fashion, leads to the resolution of the 
mentioned above problems. So, the aim of this section is to 
demonstrate first a mechanism where the four index field strength 
condensate provides  nontrivial dynamics while leaving more 
sophisticated improvements of these ideas to later sections. 

\bigskip
\subsection{Scalar field and the cosmological model of the very early 
universe}
 \bigskip

As it follows from our analysis above (see Sec.III), a model with only a
scalar field although solves the cosmological constant problem, it
cannot give rise to inflation since the gravitational effects of the scalar
field potential is always canceled by the integration constant $M$. We will
see  that  nontrivial dynamics of a single scalar field including the
possibility of inflation  can be
obtained by considering a model with an additional degree of freedom
described by a three-index potential $A_{\beta\mu\nu}$  as in the Lagrangian
density
\begin{equation}
L=-\frac{1}{\kappa}R(\Gamma,g)
+\frac{1}{2}\varphi,_{\alpha}\varphi^{,\alpha}-V(\varphi)+
\frac{1}{4!}F_{\alpha\beta\mu\nu}F^{\alpha\beta\mu\nu}.
 \label{IV1}
\end{equation}

Here 
\begin{equation}
F_{\alpha\beta\mu\nu}\equiv\partial_{[\alpha}A_{\beta\mu\nu ]}
\label{IV2}
\end{equation}
 is
the field strength which is invariant under the gauge transformation
\begin{equation}
A_{\beta\mu\nu}\rightarrow A_{\beta\mu\nu}+\partial_{[\beta}f_{\mu\nu]}
\label{IV3}
\end{equation}

In ordinary 4-dimensional GR, the
$F_{\alpha\beta\mu\nu}F^{\alpha\beta\mu\nu}$ term gives rise
to a  cosmological constant depending on an integration constant
\cite{FF1},\cite{FF2}. In our case,
due to the constraint (\ref{II3}), the degrees of freedom contained in
$F_{\alpha\beta\mu\nu}$ and those of the scalar field $\varphi$ will be
intimately correlated. The sign in front of the
$F_{\alpha\beta\mu\nu}F^{\alpha\beta\mu\nu}$ term is chosen so that in
this model the
resulting expression for the energy density of the scalar field $\varphi$ is
a positive definite one for any possible
space-time dependence of $\varphi$ in an effective "Einstein picture" (see
below). Notice also
that two last terms in the action with the Lagrangian (\ref{IV1}) break
explicitly the LES.

The gravitational equations (\ref{II2}) take now the form
\begin{equation}
-\frac{1}{\kappa}R_{\mu\nu}(\Gamma)
+\frac{1}{2}\varphi,_{\mu}\varphi,_{\nu}+
\frac{1}{6}F_{\mu\alpha\beta\gamma}F_{\nu}^{\alpha\beta\gamma}=0.
 \label{IV4}
 \end{equation}
Notice that the scalar field potential $V(\varphi)$ does not appear
explicitly in Eqs. (\ref{IV4}). However, the constraint (\ref{II3}), which
takes now  the form
\begin{equation}
V(\varphi)+M=
-\frac{1}{8}F_{\alpha\beta\mu\nu}F^{\alpha\beta\mu\nu},
 \label{IV5}
 \end{equation}
allows us to express the last term in (\ref{IV4}) in terms of the
potential $V(\varphi)$ (using the fact that
$F^{\alpha\beta\mu\nu}\propto\varepsilon^{\alpha\beta\mu\nu}$ in
4-dimensional space-time).

Varying the action with respect to $A_{\nu\alpha\beta}$, we get the equation
\begin{equation}
\partial_{\mu}(\Phi F^{\mu\nu\alpha\beta})=0
\label{IV6}
 \end{equation}
 Its general solution has
a form
\begin{equation}
F^{\alpha\beta\mu\nu}=
\frac{\lambda}{\Phi}\varepsilon^{\alpha\beta\mu\nu}
\equiv\frac{\lambda}{\chi\sqrt{-g}}\varepsilon^{\alpha\beta\mu\nu} ,
 \label{IV7}
 \end{equation}
where $\lambda$ is an integration constant. Then
$F_{\alpha\beta\mu\nu}F^{\alpha\beta\mu\nu}=-\lambda^{2}4!/\chi^{2}$ 
is not a constant now as
opposed to the GR case \cite{FF1},\cite{FF2}  and
therefore
\begin{equation}
V(\varphi)+M=3\lambda^{2}/\chi^{2}
 \label{IV8}
 \end{equation}
and
\begin{equation}
F_{\mu\alpha\beta\gamma}F_{\nu}^{\alpha\beta\gamma}=
-(6\lambda^{2}/\chi^{2})g_{\mu\nu}=-2[V(\varphi)+M]g_{\mu\nu}
\label{IV9}
 \end{equation}
This shows how
the potential $V(\varphi)$ appears in Eq. (\ref{IV4}), spontaneously 
violating
the symmetry of the action $V(\varphi)\rightarrow V(\varphi)+constant$, which
now corresponds to a redefinition of the integration constant $M$.

The equation of motion of the scalar field 
$\varphi$ is
\begin{equation}
(-g)^{-1/2}\partial_{\mu}(\sqrt{-g}g^{\mu\nu}\partial_{\nu}\varphi)
+\sigma,_{\mu}\varphi ^{,\mu}+V^{\prime}(\varphi)=0,
\label{IV10}
 \end{equation}
 where
$V^{\prime}\equiv\frac{dV}{d\varphi}$.

The derivatives of the field
$\sigma$ enter both in the gravitational Eqs. (\ref{IV4})
(through the connection) and in the scalar field equation (\ref{IV10}). In 
order to
see easily the physical content of this model, we have to perform a
{\em conformal transformation} 
\begin{equation}
\overline{g}_{\mu\nu}(x)=\chi
g_{\mu\nu}(x); \qquad \varphi\rightarrow\varphi
\label{IV11}
 \end{equation}
 to obtain an "Einstein picture".
Notice that now this transformation is not a symmetry and indeed
changes the form of equations.  In
this new frame, the gravitational equations become
\begin{equation}
G_{\mu\nu}(\overline{g}_{\alpha\beta})=\frac{\kappa}{2}T_{\mu\nu}^{eff}(\varphi)
 \label{IV12}
 \end{equation}
where
\begin{equation}
G_{\mu\nu}(\overline{g}_{\alpha\beta})=
R_{\mu\nu}(\overline{g}_{\alpha\beta})-
\frac{1}{2}\overline{g}_{\mu\nu}R(\overline{g}_{\alpha\beta})
 \label{IV13}
 \end{equation}
is the Einstein tensor in the Riemannian space-time with metric 
$\overline{g}_{\mu\nu}$, and the source is the minimally coupled scalar 
field $\varphi$
 \begin{equation}
T_{\mu\nu}^{eff}(\varphi)=
\varphi,_{\mu}\varphi,_{\nu}-
\frac{1}{2}\overline{g}_{\mu\nu}\varphi,_{\alpha}\varphi,^{\alpha}
+\overline{g}_{\mu\nu}V_{eff}(\varphi)
 \label{IV14}
 \end{equation}
with the new effective potential
\begin{equation}
V_{eff}=\frac{2}{3}\chi^{-1}(V+M)
 \label{IV15}
 \end{equation}

The scalar field equation (\ref{IV10}) in the Einstein picture takes a form
\begin{equation}
\frac{1}{\sqrt{-\overline{g}}}
\partial_{\mu}(\sqrt{-\overline{g}}\overline{g}^{\mu\nu}
\partial_{\nu}\varphi)+
\chi^{-1}V^{\prime}(\varphi)=0.
 \label{IV16}
 \end{equation}

For the possible expression for $\chi^{-1}$ we have from Eq. (\ref{IV8}) 
\begin{equation}
\frac{1}{\chi}=\pm\frac{1}{\lambda\sqrt{3}}(V+M)^{1/2}
 \label{IV17}
 \end{equation}

Independently of the sign chosen for $\chi$ in Eq. (\ref{IV17}), the 
gravitational and scalar field equations have the 
same physical content expressed in different space-time signatures. In 
what follows we simply take the + sign in (\ref{IV17}). Therefore, the 
effective scalar field potential (\ref{IV15}) has the form
\begin{equation}
V_{eff}(\varphi)=\frac{2}{\lambda 3\sqrt{3}}(V+M)^{3/2}
 \label{IV18}
 \end{equation}
and the scalar field Eq. (\ref{IV16}) becomes a conventional general 
relativistic scalar field equation with the potential $V_{eff}(\varphi)$:
\begin{equation}
\frac{1}{\sqrt{-\overline{g}}}
\partial_{\mu}(\sqrt{-\overline{g}}\quad\overline{g}^{\mu\nu}
\partial_{\nu}\varphi)+
V_{eff}^{\prime}(\varphi)=0.
 \label{IV19}
 \end{equation}

We see that in the Einstein picture, for {\em any analytic} $V(\varphi)$, \
$V_{eff}(\varphi)$ has an
extremum, that is $V^{\prime}_{eff}=0$, {\em either} when $V^{\prime}=0$
{\em or\/} $V+M=0$. The extremum when $V+M=0$ corresponds to an absolute
minimum (since $V_{eff}(\varphi)$ is non negative) and therefore it is {\em a
vacuum with
zero effective cosmological constant\/}. It should be emphasized that all
what is required is that $V+M$ touches zero at {\em some\/} point
$\varphi_{0}$ but $V^{\prime}$ at this point does not  need to vanish.
Therefore {\em no fine tuning\/} in the usual sense, of adjusting a
minimum of  a potential to coincide with the point where this
potential itself vanishes, is required. And the situation is even more
favorable since even if $V+M$ happens not to touch  zero for any value of
$\varphi$, we  always have an infinite set of other values of the 
integration constant $M$ where this will happen.

In the context of the cosmology, for the Friedmann-
Robertson-Walker universe where in the Einstein picture $\varphi
=\varphi(t)$ and 
\begin{equation}
d\overline{s}^{2}=\overline{g}_{\mu\nu}dx^{\mu}dx^{\nu}=
dt^{2}-\overline{a}^{2}(t)dl^{2},
dl^{2}=[dr^{2}/(1-kr^{2})+r^{2}d\Omega^{2}], 
\label{IV20}
 \end{equation}
we notice that due
to the positivity  of $V_{eff}$ constructed from an arbitrary analytic 
$V(\varphi)$ according to Eq. (\ref{IV18}), {\em most of the known 
inflationary 
scenarios} \cite{Gut} {\em for the very early universe can be implemented
depending on the choice of the potential $V(\varphi)$}. It is very
interesting that the parameters ruling the inflation are controlled by
the integration constants $M$ and $\lambda$.

After inflation, when the scalar field $\varphi$ approaches the position
$\varphi_{0}$ of the absolute minimum of the potential $V_{eff}$, i.e. 
when $V(\varphi)+M\rightarrow 0$, the
$\chi$-field approaches infinity as it seen from the constraint (\ref{IV8}).
To clarify the meaning of this effect, let us go back to the picture with
the original $g_{\mu\nu}$ while still using the cosmic time $t$ that was
defined in the Einstein picture. Then equation for $\varphi$ is
\begin{equation}
\ddot \varphi +3\frac{\dot a}{a}\dot \varphi-
\frac{3V^{\prime}}{4(V+M)}\dot \varphi^{2}+
\frac{\sqrt{V+M}}{\lambda\sqrt{3}}V^{\prime}=0,
 \label{IV21}
 \end{equation}
where $a^{2}(t)=\overline{a}^{2}(t)/\chi(t)$, $g_{00}(t)=1/\chi (t)$ and
constraint (\ref{IV8}) have been used.

Generally, $\dot{\varphi}$ does not go to zero as 
$\varphi\rightarrow\varphi_{0}$ (and
$V(\varphi)+M\rightarrow 0$).
In this asymptotical region we can find the first integral of Eq. 
(\ref{IV21}).
 Assuming
that $V^{\prime}(\varphi_{0})\neq 0$, i.e. without fine tuning, we get
\begin{equation}
\dot{\varphi}a^{3}(t)\simeq c[V(\varphi)+M]^{3/4},\quad c=const \quad 
(as\quad \varphi\rightarrow\varphi_{0}),
 \label{IV22}
 \end{equation}
which means that $a(t)\rightarrow 0$ as $\varphi\rightarrow\varphi_{0}$
(notice that if we would have  chosen a coordinate frame in
the original picture such that
$ds^{2}=
dt^{\prime 2}-a^{2}(t^{\prime})dl^{2}$, then instead of (\ref{IV22}) we
would have gotten $a^{3}(t^{\prime})d\varphi/dt^{\prime}\simeq
c[V(\varphi)+M]^{1/2}$ as
$\varphi\rightarrow\varphi_{0}$). Then integrating the gravitational
equations we get asymptotically (as $\varphi\rightarrow\varphi_{0}$)
that $a^{2}(t)=a_{0}^{2}/\chi(t)$, \quad $a_{0}=const$, that is in the
original frame
there is a collapse of the universe from a finite $a$ to $a=0$ in a
finite time and therefore the Riemannian curvature goes to infinity  as
$\varphi\rightarrow\varphi_{0}$.
This pathology {\em is not seen} in the
Einstein frame due to the singularity of the conformal transformation
$\overline{a}^{2}=\chi a^{2}$ at $\varphi =\varphi_{0}$. In fact, this is 
not a problem from the point of view of physics, since as
$\varphi\rightarrow\varphi_{0}$  (and $V(\varphi)+M\rightarrow 0$), the LES
becomes restored at the critical point $\varphi\equiv\varphi_{0}$ where
$V(\varphi_{0})+M=0$. In the presence of the LES, the conformal
transformation  $\overline{g}_{\mu\nu}(x)=\chi g_{\mu\nu}(x)$
becomes part of the LES transformation and represents a nonsingular gauge
choice for the metric $\overline{g}_{\mu\nu}$.

In contrast to this, there is a real problem in the scenario discussed 
above. Although the point $\varphi_{0}$ where $V(\varphi)+M=0$ satisfies 
$V^{\prime}_{eff}(\varphi_{0})=0$, \quad $V_{eff}(\varphi_{0})=0$ 
and furthermore it is an absolute minimum 
of the effective potential $V_{eff}(\varphi)$, we can see however, that the 
second derivative of $V_{eff}(\varphi)$ is equal to infinity at this point 
$\varphi_{0}$. It means that there are no physical excitations of the 
scalar field $\varphi$ around this minimum. This causes problems both in the 
cosmological picture when considering the possibility of small oscillations 
around the minimum and in the associated particle physics, since the mass 
of a scalar, like for example, the Higgs mass, will appear infinite. In 
Sec. V we will show how this problem is solved. 

\bigskip
\subsection{Gauge fields and the Higgs mechanism in the NGVE theory}

\bigskip

Let us consider now a model including gravity (formulated in the first 
order formalism), four index field strength $F_{\alpha\beta\mu\nu}$, a 
gauge field $\tilde{A}_{\mu}$ and a complex scalar field $\phi$ minimally 
coupled to the gauge field with the action
\begin{eqnarray}
S=\int\Phi d^{4}x[-\frac{1}{\kappa}R(\Gamma,g)
+\frac{1}{4!}F_{\alpha\beta\mu\nu}F^{\alpha\beta\mu\nu}
+\frac{1}{m^{4}}(\tilde{F}_{\mu\nu}\tilde{F}^{\mu\nu})^{2}
\nonumber\\
+g^{\mu\nu}(\partial_{\mu}-i\tilde{e}\tilde{A}_{\mu})\phi
           (\partial_{\nu}+i\tilde{e}\tilde{A}_{\nu})\phi^\ast
-V(|\phi|)]
 \label{IV23}
\end{eqnarray}
where $\tilde{F}_{\mu\nu}\equiv\partial_{\mu}\tilde{A}_{\nu}-
\partial_{\nu}\tilde{A}_{\mu}$.

Notice that the kinetic term of the gauge field $\tilde{A}_{\mu}$ is 
chosen in an unusual way where an additional parameter $m$ with 
dimensionality of mass is introduced to provide the canonical 
dimensionality for the gauge field $\tilde{A}_{\mu}$. The reason for such 
a choice of the kinetic term of the gauge field is to achieve for it the 
same degree of homogeneity in $g^{\mu\nu}$ in the Lagrangian density as we 
have for the four index field strength $F_{\alpha\beta\mu\nu}$. As we 
will see, for such a choice, after solving the constraint we obtain the 
standard effective low energy physics. For example, in the absence of 
other interactions, the gauge field equations possess conformal 
invariance or, what is the same, they have the standard Maxwell form.

By making use the gauge invariance we choose the unitary gauge (where 
$Im\,\phi (x)=0$) and then the Lagrangian density takes the form
\begin{eqnarray}
L=-\frac{1}{\kappa}R(\Gamma,g)
+\frac{1}{4!}F_{\alpha\beta\mu\nu}F^{\alpha\beta\mu\nu}
+\frac{1}{m^{4}}(\tilde{F}_{\mu\nu}\tilde{F}^{\mu\nu})^{2}+
\nonumber\\
\frac{1}{2}g^{\mu\nu}\varphi,_{\mu}\varphi,_{\nu}-
V(\varphi)+
\frac{1}{2}\tilde{e}^{2}\varphi^{2}g^{\mu\nu}\tilde{A}_{\mu}\tilde{A}_{\nu})
 \label{IV24}
\end{eqnarray}
where we have defined $|\phi| =\frac{1}{\sqrt{2}}\varphi$.

The constraint (\ref{II3}) corresponding to the Lagrangian density 
(\ref{IV24}) is
\begin{equation}
\frac{1}{8}F_{\alpha\beta\mu\nu}F^{\alpha\beta\mu\nu}
+\frac{3}{m^{2}}(\tilde{F}_{\mu\nu}\tilde{F}^{\mu\nu})^{2}+
V(\varphi)+M=0
 \label{IV25}
\end{equation}
Similar to the first and the fourth terms, the last term 
in Eq. (\ref{IV24}) does not contribute to the constraint since it is 
homogeneous of degree one in $g^{\mu\nu}$.

The equation for $F_{\alpha\beta\mu\nu}$ and its solution are still the 
same as in Eqs. (\ref{IV6}), (\ref{IV7}) which bring constraint 
(\ref{IV25}) to the following equation:
\begin{equation}
\frac{\omega^{2}m^{4}}{\chi^{2}}=
\frac{1}{m^{4}}(\tilde{F}_{\mu\nu}\tilde{F}^{\mu\nu})^{2}+
\frac{1}{3}(V(\varphi)+M)
 \label{IV26}
\end{equation}
where we have defined $\lambda$ in terms of the mass parameter $m^{2}$ 
\begin{equation}
\lambda=\omega m^{2}
 \label{IV27}
\end{equation}
$\omega$ being a dimensionless constant.

Varying the action with respect to $\tilde{A}_{\mu}$ we get 
\begin{equation}
\frac{1}{\sqrt{-g}}\partial_{\mu}[\sqrt{-g}\chi 
(\tilde{F}_{\alpha\beta}\tilde{F}^{\alpha\beta})g^{\mu\gamma}g^{\nu\delta}
\tilde{F}_{\gamma\delta}]
-\frac{\tilde{e}^{2}m^{4}}{8}\varphi^{2}\chi g^{\alpha\nu}\tilde{A}_{\alpha}
=0
 \label{IV28}
\end{equation}

Looking at gauge field fluctuations around the true vacuum 
$\varphi=\varphi_{0}$ where $V(\varphi_{0})+M=0$ and 
$V^{\prime}_{eff}(\varphi_{0})=0$ 
(and ignoring the scalar field fluctuations around the true vacuum
$\varphi_{0}$,  see the previous 
subsection), we get from  (\ref{IV26})
\begin{equation}
\chi (\tilde{F}_{\mu\nu}\tilde{F}^{\mu\nu})=
\pm\omega m^{4}
 \label{IV29}
\end{equation}
First of all notice that in the case where is no coupling to the scalar 
field ($\tilde{e}=0$), Eq. (\ref{IV28}) after making use Eq. (\ref{IV29}) 
becomes the Maxwell's equations 
\begin{equation}
\frac{1}{\sqrt{-g}}\partial_{\mu}({\sqrt{-g}}
g^{\mu\gamma}g^{\nu\delta}
\tilde{F}_{\gamma\delta})
=0
 \label{IV30}
\end{equation}
which are indeed conformally invariant.

In the presence of interactions with scalar field ($\tilde{e}\not=0$), 
Eqs. (\ref{IV28}) and (\ref{IV29}) lead to the singularity in the second 
term when $\tilde{F}_{\alpha\beta}\tilde{F}^{\alpha\beta}=0$.

It is worthwhile to remind that in the context of cosmology we studied 
in Subsection IVA, the $\chi$-field becomes divergent as $\varphi$ 
approaches the absolute minimum $\varphi_{0}$. Therefore it is not a 
surprise that a singularity also occurs in the case where 
$\tilde{F}_{\alpha\beta}\tilde{F}^{\alpha\beta}=0$, that is when electric 
dominated field evolves into magnetic dominated one (or vice versa). 
Similar to what happens in the cosmological scenario, here we will see 
that this singularity is eliminated by the same conformal transformation 
(\ref{IV11}), that is by transforming to the Einstein frame.

In fact, performing the conformal transformation (\ref{IV11}) and taking 
into account the constraint (\ref{IV29}) we obtain for the gauge field 
equation in the Einstein frame
\begin{equation}
\frac{1}{\sqrt{-\overline{g}}}\partial_{\mu}[\pm\sqrt{-\overline{g}}
\enspace\overline{g}^{\mu\gamma}\overline{g}^{\nu\delta}
\tilde{F}_{\gamma\delta}]
-\frac{\tilde{e}^{2}}{8\omega}\varphi^{2}\overline{g}^{\alpha\nu}
\tilde{A}_{\alpha}
=0
 \label{IV31}
\end{equation}

Notice that we have not taken the sign $\pm$ in Eq. (\ref{IV31}) outside 
the derivative operator. This makes it apparent that if we pick one of 
the two branches displayed in Eq. (\ref{IV31}), it cannot evolve 
continuously to the another branch. Therefore the requirement of the 
analyticity of the resulting equations exclude for example the alternative
$\chi |\tilde{F}_{\mu\nu}\tilde{F}^{\mu\nu}|=
\omega m^{4}$ instead of (\ref{IV29}) as a solution of the constraint 
(\ref{IV26}) at the point $\varphi=\varphi_{0}$.

When choosing one of the branches in Eq. (\ref{IV29}) we note that the 
conformal transformation (\ref{IV11}) changes the relative signatures of 
the original metric $g_{\mu\nu}$ and the metric $\overline{g}_{\mu\nu}$ in 
the Einstein frame when the gauge field evolves from electric dominated 
to magnetic dominated (and vice versa).

Similar to what happens in the case of cosmology (see Subsection IVA), Eq. 
(\ref{IV31}) shows that in the Einstein frame there is no singularity 
when $\chi^{-1}\rightarrow\pm 0$. Therefore, changes of the signature of 
the metric take place only in the original frame and not in the Einstein 
frame.

For the choice of the signature $(+---)$ in the Einstein frame we have to 
choose the branch $(-)$ in Eq.
(\ref{IV31}) in order to avoid tachyonic behavior. After the change of 
notations
\begin{equation}
e=\frac{\tilde{e}}{2\sqrt{2\omega}};\qquad 
A_{\mu}=2\sqrt{2\omega}\tilde{A}_{\mu}; 
 \label{IV32}
\end{equation} 
we get the canonical form of equations for the vector field 
\begin{equation}
\frac{1}{\sqrt{-\overline{g}}}\partial_{\mu}(\sqrt{-\overline{g}}
\enspace\overline{g}^{\mu\gamma}\overline{g}^{\nu\delta}
F_{\gamma\delta})
+m_{A}^{2}A^{\nu}
=0
 \label{IV33}
\end{equation}
 where
\begin{equation}
m_{A}^{2}=e^{2}\varphi_{0}^{2}
 \label{IV34}
\end{equation}
is the mass of the vector boson $A_{\mu}$ which is generated by the 
spontaneous symmetry breaking (SSB) of the gauge invariance when the scalar 
field $\phi$ is located at the absolute minimum $|\phi|=
\frac{1}{\sqrt{2}}\varphi =\frac{1}{\sqrt{2}}\varphi_{0}$ of the effevtive 
potential (\ref{IV18}). We have used the notations
\begin{equation}
A^{\mu}=\overline{g}^{\mu\nu}A_{\nu}; \ F_{\mu\nu}=\partial_{\mu}A_{\nu}-
\partial_{\nu}A_{\mu}; \ 
F^{\mu\nu}=\overline{g}^{\mu\alpha}\overline{g}^{\nu\beta}F_{\alpha\beta}
 \label{IV35}
\end{equation} 

The appropriate gravitational equations at the absolute minimum 
$\varphi_{0}$ in the Einstein frame, when the 
source is the vector field $A_{\mu}$ takes the standard form
\begin{equation}
G_{\mu\nu}(\overline{g}_{\alpha\beta})=\frac{\kappa}{2}T_{\mu\nu}
(A_{\alpha})
 \label{IV36}
\end{equation}
where
\begin{equation}
T_{\mu\nu}(A_{\alpha})=
\frac{1}{4}\overline{g}_{\mu\nu}F_{\alpha\beta}F^{\alpha\beta}
-F_{\mu\alpha}F_{\nu\beta}\overline{g}^{\alpha\beta}+
\frac{1}{2}m_{A}^{2}(A_{\mu}A_{\nu}-
\frac{1}{2}\overline{g}_{\mu\nu}A_{\alpha}A^{\alpha})
 \label{IV37}
\end{equation} 

For the scalar field equation in the Einstein frame we obtain 
\begin{equation}
\frac{1}{\sqrt{-\overline{g}}}\partial_{\mu}(\sqrt{-\overline{g}}
\enspace\overline{g}^{\mu\nu}\partial_{\nu}\varphi)-
e^{2}A_{\alpha}A^{\alpha}\varphi+V^{\prime}_{eff}=0
 \label{IV38}
\end{equation}
with $V_{eff}$ given by Eq. (\ref{IV18}).

As we see, the rescaling (\ref{IV32}) provides the canonical 
normalization of the gauge field so to reproduce the standard form of the 
energy-momentum tensor (\ref{IV37}) and standard interaction of the 
vector field $A_{\mu}$ to the scalar field after SSB. A very interesting 
feature of the theory in the Einstein picture is the fact that the gauge 
coupling constant $e$  depends on the integration constant $\omega$ which 
appears in the solution for the four index field condensate (\ref{IV7}) 
(see also the definition (\ref{IV27}).

Finally we have to notice that it is possible to improve the model 
discussed in this Sec.IV (see for example \cite{GK4}) in such a way that 
the mass of the scalar field becomes finite (compare with discussion at 
the end of Subsec.IVA). Then, however, at the very high energies where 
the scalar field fluctuations around the vacuum $\varphi=\varphi_{0}$
have to be taken into account, it follows from the constraint 
(\ref{IV26}) and Eq. (\ref{IV28}) that the nonminimal nonrenormalizable 
interaction of the gauge field with the scalar field appears.

\bigskip
\section{Unified gauge sector and models with realistic particle 
fields dynamics and cosmology}

\bigskip

As we have seen in Sec. IV, it is possible to realize a nontrivial 
dynamics of a scalar field and a gauge field while solving the 
cosmological constant problem. However, the simplest models presented in 
Sec. IV give rise to nonrenormalizable coupling between gauge and scalar 
fields. We are going to show in this section that by a certain sort of 
{\em unification of all gauge fields} we can get rid of the above 
mentioned problem.

The basic idea consists of demanding that the dependence of the 
Lagrangian density on the gradients of the gauge field potentials is only 
through a single variable which is the sum of all possible kinetic terms 
and a corresponding four index field strength term. The fact that all 
gauge fields must come together is automatic in a unified gauge models 
where the Lagrangian density must depend only on 
$F_{\mu\nu}^{a}F^{a\mu\nu}$ where $a$ is for example an $SU(5)$ index. In 
addition we insist also in introducing the three index potential 
$A_{\mu\nu\alpha}$ into the game in a similar way. Here we will demand 
that the field strength $\partial_{[\alpha}A_{\mu\nu\beta ]}$ is combined 
with other usual gauge field kinetic terms, like 
$F_{\mu\nu}^{a}F^{a\mu\nu}$, in such a way that the homogeneity of each 
of the terms in $g^{\mu\nu}$ is of degree 2. This singles out the 
following combination analytic in the gradients of the gauge fields 
potentials 
\footnote {The other possibility including
$\sqrt{-F_{\mu\nu\alpha\beta}F^{\mu\nu\alpha\beta}}\equiv
|\frac{\varepsilon^{\mu\nu\alpha\beta}}{\sqrt{-g}}
\partial_{\mu}A_{\nu\alpha\beta}|$ is not analytic one.}
\begin{equation}
y\equiv F_{\mu\nu}F^{\mu\nu}+m^{2}\frac{\varepsilon^{\mu\nu\alpha\beta}} 
{\sqrt{-g}} \partial_{\mu}A_{\nu\alpha\beta}
 \label{V1}
\end{equation}
We will call $y$ the gauge complex. Here  $m$ is a parameter with 
the dimensions of mass.

The demand that the term which depends on the condensate of 
$A_{\mu\nu\alpha}$, has to have the same transformation under 
$g^{\mu\nu}\rightarrow \Omega g^{\mu\nu}$ as the ordinary gauge fields, 
finds a simple analogy in a related higher dimensional picture where the 
components of the gauge field strengths in the direction of the extra 
dimensions can play a similar role to that of the $A_{\mu\nu\alpha}$ 
field in 4-dimensional space-time. In 6-dimensional case for example, 
$F_{AB}F^{AB}\equiv 
F_{\mu\nu}F^{\mu\nu}+2F_{a\mu}F^{a\mu}+F_{ab}F^{ab}$, where $\mu,\nu 
=0,1,2,3$; \ $a,b$=4,5 and $F_{ab}F^{ab}$ plays then the role of 
$\frac{\varepsilon^{\mu\nu\alpha\beta}}{\sqrt{-g}}
\partial_{\mu}A_{\nu\alpha\beta}$  in the condensate state \cite{G} (here 
$F_{ab}$ takes a "magnetic monopole" expectation value). Then the 
requirement of equal behavior under conformal transformation is of course 
automatic.

Another independent problem we have discussed in Sec. IVA, is the 
infinite value of masses of scalar fields. Concerning to the resolution 
of this problem we have device two approaches: a) The first one (see 
Subsection VA) uses a critical limit of a family of Lagrangians. In this 
limit the scalar field acquires a finite mass \cite{GK4}. b) The second 
approach (see
Subsection VB), which seems more generic, is based on the appearance of a 
"persistent condensate".

\bigskip
\subsection{Model with a critical limit}

\bigskip   

Following the guidelines described above we consider Lagrangians which 
depend on the ordinary gauge field $\tilde{A}_{\mu}$ and the three index 
gauge field $A_{\mu\nu\alpha}$ only through the gauge complex $y$, 
Eq.(\ref{V1}). At first we consider the interesting family of 
Lagrangians with a power low dependence on $y$. Therefore, in the 
unitary gauge for $\tilde{A}_{\mu}$, instead of dealing with 
(\ref{IV23}),(\ref{IV24}), we have the action 
\begin{eqnarray}
S=\int\Phi d^{4}x\left[-\frac{1}{\kappa}R(\Gamma,g)
-\frac{1}{pm^{4(p-1)}}y^{p}+
\frac{1}{2}g^{\mu\nu}\varphi,_{\mu}\varphi,_{\nu}-
V(\varphi)+
\frac{1}{2}\tilde{e}^{2}\varphi^{2}g^{\mu\nu}\tilde{A}_{\mu}\tilde{A}_{\nu}
\right]
 \label{V2}
\end{eqnarray}
where dimensionless 
parameter $p$ is a real number. As we will see later, the physically 
interesting case is achieved in the critical limit $p\rightarrow\infty$.

Constraint (\ref{II3}) takes now the form
\begin{equation}
\frac{2p-1}{pm^{4(p-1)}}y^{p}=V(\varphi)+M
 \label{V3}
\end{equation}
which defines $y$ as a function of $\varphi$.

From variation with respect to $A_{\nu\alpha\beta}$ we obtain the 
equation 
\begin{equation}
\partial_{\mu}(\chi y^{p-1}\varepsilon^{\mu\nu\alpha\beta})=0
 \label{V4}
\end{equation}
the solution of which can be written as
\begin{equation}
\chi y^{p-1}=\omega m^{4(p-1)}
 \label{V5}
\end{equation}
where $\omega$ is a dimensionless integration constant.

Varying with respect to the scalar field $\varphi$ we get
\begin{equation}
(-g)^{-1/2}\partial_{\mu}(\sqrt{-g}g^{\mu\nu}\partial_{\nu}\varphi)
+\sigma,_{\mu}\varphi ^{,\mu}+V^{\prime}(\varphi)+
\tilde{e}^{2}\varphi g^{\alpha\beta}\tilde{A}_{\alpha}\tilde{A}_{\beta}=0,
 \label{V6}
\end{equation}

The equation for the gauge field $\tilde{A}_{\alpha}$ is
\begin{equation}
\frac{1}{\sqrt{-g}}\partial_{\mu}[\chi y^{p-1}\sqrt{-g}
\tilde{F}^{\mu\nu}]
+\frac{\tilde{e}^{2}}{4}m^{4(p-1)}\varphi^{2}\chi\tilde{A}^{\nu}
=0
 \label{V7}
\end{equation}
which becomes
\begin{equation}
\frac{1}{\sqrt{-g}}\partial_{\mu}[\sqrt{-g}\tilde{F}^{\mu\nu}]
+\frac{\tilde{e}^{2}}{4\omega}\varphi^{2}\chi\tilde{A}_{\nu}
=0
 \label{V8}
\end{equation}
due to Eq. (\ref{V5}). And finally, the variation of $g^{\mu\nu}$ leads 
to the gravitational equations
\begin{eqnarray}
\frac{1}{\kappa}R_{\mu\nu}(\Gamma,g)=
-\frac{y^{p}}{2m^{4(p-1)}}g_{\mu\nu}+
\frac{y^{p-1}}{m^{4(p-1)}}\left[\frac{1}{2}\tilde{F}^{\alpha\beta}
\tilde{F}_{\alpha\beta}g_{\mu\nu}-
2\tilde{F}_{\mu\alpha}\tilde{F}_{\nu\beta}g^{\alpha\beta}\right]+
\frac{1}{2}\varphi_{,\mu}\varphi_{,\nu}+
\frac{\tilde{e}^{2}}{2}\varphi^{2}\tilde{A}_{\mu}\tilde{A}_{\nu}
 \label{V9}
\end{eqnarray}
where Eq. (\ref{V1}) has been used.

For the same reasons as those explained in Sec. IV, we have to perform 
the conformal transformation (\ref{IV11}) which provides a formulation of 
the theory in the Einstein picture.  Eqs. 
(\ref{V6}), (\ref{V8}) and (\ref{V9}) become then correspondingly
\begin{equation}
\frac{1}{\sqrt{-\overline{g}}}\partial_{\mu}(\sqrt{-\overline{g}}
\enspace\overline{g}^{\mu\nu}\partial_{\nu}\varphi)+
\frac{dV^{(p)}_{eff}}{d\varphi}+
e^{2}\varphi g^{\alpha\beta}A_{\alpha}A_{\beta}
=0
 \label{V10}
\end{equation}
\begin{equation}
\frac{1}{\sqrt{-\overline{g}}}\partial_{\mu}(\sqrt{-\overline{g}}
\enspace\overline{g}^{\mu\alpha}\overline{g}^{\nu\beta}
F_{\alpha\beta})+
e^{2}\varphi^{2}\overline{g}^{\nu\alpha}A_{\alpha}
=0
 \label{V11}
\end{equation} 
\begin{equation}
G_{\mu\nu}(\overline{g}_{\alpha\beta})=\frac{\kappa}{2}T_{\mu\nu}
 \label{V12}
\end{equation}
\begin{equation}
T_{\mu\nu}=
\varphi_{,\mu}\varphi_{,\nu}-
\frac{1}{2}g_{\mu\nu}\varphi_{,\alpha}\varphi_{,\beta}
\overline{g}^{\alpha\beta}+V^{(p)}_{eff}(\varphi)\overline{g}_{\mu\nu}+
\frac{1}{4}\overline{g}_{\mu\nu}F_{\alpha\beta}F^{\alpha\beta}
-F_{\mu\alpha}F_{\nu\beta}\overline{g}^{\alpha\beta}+
e^{2}\varphi^{2}(A_{\mu}A_{\nu}-
\frac{1}{2}\overline{g}_{\mu\nu}A_{\alpha}A^{\alpha})
 \label{V13}
\end{equation}
where
\begin{equation}
V^{(p)}_{eff}(\varphi)\equiv\left[\omega m^{4(1-1/p)}\right]^{-1}
\left(\frac{p}{2p-1}\right)^{2-1/p}(V(\varphi)+M)^{2-1/p}
 \label{V14}
\end{equation}
and the rescalings
\begin{equation}
A_{\mu}\equiv 2\sqrt{\omega}\tilde{A}_{\mu}; \qquad 
e=\frac{\tilde{e}}{2\sqrt{\omega}}
 \label{V15}
\end{equation}
have been performed. It is assumed that $\omega>0$. 

Equations (\ref{V10})-(\ref{V13}) describe the family of canonical 
equations of GR (parametrized by the parameter $p$) for a gauge model (in 
the unitary gauge) 
including gauge field $A_{\mu}$ minimally coupled to the scalar field
$\phi$ with the potential (\ref{V14}) and the coupling constant $e$. For 
$p=1$ (which has to be studied by itself), the theory reproduces the 
Einstein GR with the original potential $V(\varphi)$ and with a 
cosmological constant $M$ (see Ref. \cite{GK2}). In contrast, for any 
$p>1$, the effective potential $V^{(p)}_{eff}(\varphi)$, given by 
Eq.(\ref{V14}), has an absolute minimum at the point $\varphi_{0}$ where 
$V(\varphi_{0})+M=0$, which generalizes the results of Sec. IV.

However an additional remarkable feature appears when 
$p\rightarrow\infty$. In this case
\begin{equation}
V_{eff}(\varphi)\equiv\lim_{p \to \infty}V^{(p)}_{eff}(\varphi)=
\frac{1}{4\omega m^{4}}(V+M)^{2}
 \label{V16}
\end{equation}
and for any analytical function $V(\varphi)$, all derivatives of the 
effective potential $V_{eff}(\varphi)$ 
are finite at the
absolute minimum $\varphi=\varphi_{0}$ where $V(\varphi_{0})+M=0$. In
particular, $V^{\prime\prime}_{eff}(\varphi_{0})\propto
[V^{\prime}(\varphi_{0})]^{2}$ is finite (and nonzero if we do not carry out
the fine tuning $V^{\prime}(\varphi_{0})=0$). Therefore the Higgs boson, in
particular, can reappear
as a physical particle of the theory. In the context of  cosmology
where $V_{eff}(\varphi)$ plays the role of the inflaton potential, a finite
mass of the inflaton allows to recover the usual oscillatory regime of
the reheating period after inflation that are usually considered.

Notice that again (as it was in Sec. IV) the gauge coupling constant $e$ 
in the Einstein frame (see Eqs. (\ref{V15})) depends on the integration 
constant $\omega$ which appears in the solution for the four index field 
condensate (\ref{V5}).

From Eqs.(\ref{V3}) and (\ref{V5}) it follows that
\begin{equation}
\chi =(2-\frac{1}{p})^{1-1/p}\omega m^{4(1-1/p)}(V+M)^{-1+1/p}
 \label{V17}
\end{equation}
so that if $p>1$ or $p<0$ we obtain that $\chi\rightarrow\infty$ as 
$V+M\rightarrow 0$ which generalizes the situation described in 
Subsection IVA.

It is very instructive to look at what happens to the condensate $y$ when 
we approach the true vacuum $\varphi=\varphi_{0}$ where 
$V(\varphi_{0})+M=0$. For any finite $p>1$ we see from Eq. (\ref{V3}) 
that $y=(V+M)^{1/p}m^{4-4/p}(\frac{p}{2p-1})^{1/p}$ and $y\rightarrow 0$ 
in this limit. We notice however the very interesting effect which 
consists of the fact that as $p$ becomes big, $y$ approaches zero but at 
a very slow rate. In the limit $p\rightarrow\infty$ we can indeed argue 
that $y$ does not necessarily approaches zero but rather to an 
undetermined constant since $y\sim 0^{0}$ which is not defined. This 
suggests that the possible existence of a condensate $y$ that survives 
even in the true vacuum (which we will call "persistent condensate") is 
the cause of the remarkable feature which allows $V_{eff}$ to be of the 
form (\ref{V16}). In the next subsection we will indeed verify this 
explicitly.

\bigskip
\subsection{Model with persistent condensate}

\bigskip 

To implement the suggestion discussed at the end of the previous 
subsection let us consider a model with the action (expressed in the 
unitary gauge)
\begin{eqnarray}
S=\int\Phi d^{4}x[-\frac{1}{\kappa}R(\Gamma,g)-m^{4}f(u)+
\frac{1}{2}g^{\mu\nu}\varphi,_{\mu}\varphi,_{\nu}-
V(\varphi)+
\frac{1}{2}\tilde{e}^{2}\varphi^{2}g^{\mu\nu}\tilde{A}_{\mu}\tilde{A}_{\nu}]
 \label{V18}
\end{eqnarray}
where we have used the notations of Secs. IVB and VA and $f(u)$ is a 
function of the dimensionless argument $u=y/m^{4}$. We will see that the 
only requirement condition on the function $f(u)$, that provides a 
persistent condensate with physically reasonable consequences, is that 
$f^{\prime}(u)\equiv\frac{df}{du}=0$ for some $y=y_{0}>0$.

The constraint (\ref{II3}) has now the form 
\begin{equation}
-2uf^{\prime} (u)+f(u)+\frac{1}{m^{4}}[V(\varphi)+M]=0.
 \label{V19}
\end{equation}

Varying with respect to $A_{\nu\alpha\beta}$ we get
\begin{equation}
\partial_{\mu}(\chi f^{\prime}\varepsilon^{\mu\nu\alpha\beta})=0
 \label{V20}
\end{equation}
which gives
\begin{equation}
\chi f^{\prime} =\omega
 \label{V21}
\end{equation}
where $\omega$ is a dimensionless integration constant and Eq. 
(\ref{V21}) replaces Eq. (\ref{V5}).

Equation for the gauge field $\tilde{A}_{\mu}$ 
\begin{equation}
\frac{1}{\sqrt{-g}}\partial_{\mu}[\chi f^{\prime}\sqrt{-g}
\tilde{F}^{\mu\nu}]
+\frac{\tilde{e}^{2}}{4}\varphi^{2}\chi\tilde{A}^{\nu}
=0
 \label{V22}
\end{equation}
is reduced exactly to the form of Eq. (\ref{V8}) due to Eq. (\ref{V21}).

The gravitational equations originated by the variation of $g^{\mu\nu}$ 
take the form
\begin{eqnarray}
\frac{1}{\kappa}R_{\mu\nu}(\Gamma,g)=
-\frac{1}{2}yf^{\prime}g_{\mu\nu}+
\frac{1}{2}f^{\prime}\left[\tilde{F}^{\alpha\beta}
\tilde{F}_{\alpha\beta}g_{\mu\nu}-
4\tilde{F}_{\mu\alpha}\tilde{F}_{\nu\beta}g^{\alpha\beta}\right]
+\frac{1}{2}\varphi_{,\mu}\varphi_{,\nu}+
\frac{\tilde{e}^{2}}{2}\varphi^{2}\tilde{A}_{\mu}\tilde{A}_{\nu}
 \label{V23}
\end{eqnarray}
after using Eq. (\ref{V1}).

The scalar field equation has the same form as Eq. (\ref{V6}).

Performing the conformal transformation (\ref{IV11}) to the Einstein 
frame and rescaling $\tilde{A}_{\mu}$ and $\tilde{e}^{2}$ to ${A}_{\mu}$ 
and ${e}^{2}$ by making use Eqs. (\ref{V15}) with $\omega >0$, we obtain 
equations of the form of Eqs. (\ref{V10})-(\ref{V13}) where 
$\frac{dV_{eff}}{d\varphi}$ in the scalar field equation is now
\begin{equation}
\frac{dV_{eff}}{d\varphi}\equiv V_{eff}^{\prime}=
\frac{1}{\omega}\frac{df}{du}\frac{dV}{d\varphi}
 \label{V24}
\end{equation}
and instead of $V_{eff}^{(p)}$ in the gravitational Eqs. (\ref{V13}), the 
effective scalar field potential appears of the form
\begin{equation}
 V_{eff}(\varphi)=
\frac{y}{\omega}(f^{\prime}(u))^{2}
 \label{V25}
\end{equation}
Here $y, \ u\equiv y/m^{4}$ and $f(u)$ are functions of $\varphi$ due to 
the constraint (\ref{V19}). As in the previous examples, it can be shown 
that two different form of appearance of $V_{eff}$ and $V^{\prime}_{eff}$ 
in the gravitational field equations and in the scalar field equation 
correspondingly, are selfconsistent (the reason for this is the existence 
of Bianchi identities). This consistency also may be shown by taking the 
derivative of Eq. (\ref{V25}) with respect to $\varphi$ and using the 
derivative of the constraint (\ref{V19}) with respect to $\varphi$. As the 
result we obtain Eq. (\ref{V24}). 

Looking at Eq. (\ref{V24}) we see that there are two ways of obtaining an 
extremum of $V_{eff}(\varphi)$:

 (a) The first one is when 
$\frac{dV}{d\varphi}=0$ which corresponds to an extremum of the original 
potential $V(\varphi)$. In this case there is no reason for the vanishing 
$V_{eff}(\varphi)$ in such an extremum if we do not resort to some kind 
unnatural fine tuning.

 (b) The second way is to consider the situation where 
\begin{equation}
\frac{df}{du}|_{u=u_{0}\not=0}=0
 \label{V26}
\end{equation} 
and the appropriate value of $\varphi_{0}$ is related to $u_{0}$ by the 
constraint (\ref{V19}). For this extremum of the effective potential we 
see immediately from Eq. (\ref{V25}) that $V_{eff}(\varphi_{0})=0$.

If we assume that $y_{0}=\frac{u_{0}}{m^{4}}$ is positive then it is 
clear also from  (\ref{V25}) that $\varphi_{0}$ (where 
$V_{eff}(\varphi_{0})=0$, 
$\frac{dV_{eff}}{d\varphi}|_{\varphi=\varphi_{0}}=0$), is a 
minimum since any small fluctuations bring us to a higher positive value of 
$V_{eff}$. In this case the vacuum is defined both by value of the gauge 
complex condensate $u=u_{0}$ and by the scalar condensate $\varphi 
=\varphi_{0}$ satisfying the condition
\begin{equation}
f(u_{0})+\frac{1}{m^{4}}[V(\varphi_{0})+M]=0.
 \label{V27}
\end{equation}
which follows from the constraint (\ref{V19}) and Eq.(\ref{V26}). Notice 
that transition to the Einstein frame does not change the value of the 
gauge complex condensate $u_{0}$ since it is defined by the value of 
$\varphi_{0}$ due to Eq. (\ref{V27}). 
 
It is very important to notice that Eq. (\ref{V27}) represents {\em the 
exact mutual cancellation of the contributions to the vacuum energy of 
the integration constant $M$, the scalar field condensate and 
the gauge complex condensate}.

It can be shown explicitly (by using the constraint (\ref{V19}), its 
derivative and Eq. (\ref{V24}) that the mass square of the scalar 
particle is
\begin{equation}
V^{\prime\prime}_{eff}(\varphi_{0})=
\frac{1}{2\omega y_{0}}[V^{\prime}(\varphi_{0})]^{2}.
 \label{V28}
\end{equation}  
which is positive if both $\omega >0$ and $y_{0}>0$.

\bigskip
\section{The inclusion of fermions}

\bigskip

\bigskip
\subsection{Fermions in the NGVE theory}

\bigskip

To present a complete enough picture let us consider a model including 
gravity, gauge and scalar field sectors (as in the action (\ref{V18}), 
once again in the unitary gauge) and, in addition, the fermionic sector 
(see Appendix B):
\begin{eqnarray}
S=\int\Phi d^{4}x[-\frac{1}{\kappa}V^{a\mu}V^{b\nu}R_{\mu\nu 
ab}(\omega)-m^{4}f(u) +\frac{1}{2}g^{\mu\nu}\varphi,_{\mu}\varphi,_{\nu}-
V(\varphi)+
\frac{1}{2}\tilde{e}^{2}\varphi^{2}g^{\mu\nu}\tilde{A}_{\mu}\tilde{A}_{\nu}
\nonumber\\
+\frac{i}{2}\overline{\Psi}\left\{\gamma^{a}V_{a}^{\mu}
(\overrightarrow{\partial}_{\mu}+\frac{1}{2}\omega_{\mu}^{cd}\sigma_{cd}-
i\tilde{e}\tilde{A}_{\mu})
-(\overleftarrow{\partial}_{\mu}-\frac{1}{2}\omega_{\mu}^{cd}\sigma_{cd}+
i\tilde{e}\tilde{A}_{\mu})
\gamma^{a}V_{a}^{\mu}\right\}\Psi+U(\overline{\Psi}\Psi)]
 \label{VI1}
 \end{eqnarray}
where the selfinteraction term $U(\overline{\Psi}\Psi)$ depending on the 
argument $\overline{\Psi}\Psi$ remains unspecified in this subsection.

Equation for $\Psi$ which follows from the action (\ref{VI1}) is
\begin{equation}
\left\{i\left [V_{a}^{\mu}\gamma^{a}\left 
(\partial_{\mu}-i\tilde{e}\tilde{A}_{\mu}\right )+
\gamma^{a}C^{b}_{ab}
+\frac{1}{4}\omega_{\mu}^{cd}(\gamma^{a}\sigma_{cd}+
                              \sigma_{cd}\gamma^{a})V^{\mu}_{a}\right ]+
\frac{1}{2}\gamma^{a}V^{\mu}_{a}\sigma_{,\mu}
+U^{\prime}(\overline{\Psi}\Psi)\right\}\Psi=0
 \label{VI2}
 \end{equation}
where 
\begin{equation}
C^{b}_{ab}=\frac{1}{2\sqrt{-g}}
\partial_{\mu}\left (\sqrt{-g}V^{\mu}_{a}\right )
 \label{VI3}
\end{equation}
is the trace of the so-called Ricci rotation coefficients \cite{Gasp} and 
$U^{\prime}$ is derivative of $U$ with respect to its argument 
$\overline{\Psi}\Psi$. Spin-connection $\omega_{\mu}^{cd}$ is defined by 
Eqs. (\ref{C4})- (\ref{C7}). Remind that $\sigma$-field 
is defined by Eq. (\ref{ski}).

After the transition to the Einstein frame by means of the conformal 
transformations
\begin{equation}
V^{\prime}_{a\mu}(x)=\chi^{1/2}(x)V_{a\mu}(x)
 \label{VI4}
\end{equation}
\begin{equation}
\Psi^{\prime} (x)=\chi^{-1/4}(x)\Psi (x); \ 
\overline{\Psi}^{\prime} (x)=\chi^{-1/4}(x)\overline{\Psi} (x)
 \label{VI5}
\end{equation}
\begin{equation}
\varphi\rightarrow\varphi ; \ \tilde{A}_{\mu}\rightarrow\tilde{A}_{\mu}
 \label{VI6}
\end{equation}
and performing the rescalings (\ref{V15}), Eq. (\ref{VI2}) is reduced to the 
form
\begin{equation}
\left\{i\left [V_{a}^{\prime\mu}\gamma^{a}\left
(\partial_{\mu}-ieA_{\mu}\right )+
\gamma^{a}C^{\prime b}_{ab}
+\frac{i}{4}\omega_{\mu}^{\prime cd}\varepsilon_{abcd}\gamma^{5}\gamma^{b}
V^{a\mu}\right ]
+\chi^{-1/2}U^{\prime}(\overline{\Psi}^{\prime}\Psi^{\prime})\right\}
\Psi^{\prime}=0
 \label{VI7}
 \end{equation}
where $U^{\prime}(\overline{\Psi}^{\prime}\Psi^{\prime})=
U^{\prime}(\overline{\Psi}\Psi)|_{\Psi=\chi^{1/4}\Psi^{\prime};
\overline{\Psi}=\chi^{1/4}\overline{\Psi}^{\prime}}$ and $\chi$ enters 
only as a factor in front of 
$U^{\prime}(\overline{\Psi}^{\prime}\Psi^{\prime})$. In Eq. (\ref{VI7}), 
\begin{equation}
C^{\prime b}_{ab}=\frac{1}{2\sqrt{-g^{\prime}}}
\partial_{\mu}\left (\sqrt{-g^{\prime}}V^{\prime\mu}_{a}\right )
 \label{VI8}
 \end{equation}
and the spin-connection $\omega_{\mu}^{\prime cd}$ is
\begin{equation}
\omega_{\mu}^{\prime cd}=\omega_{\mu}^{cd}(V^{\prime\alpha}_{a})+
K_{\mu}^{cd}(V^{\prime\alpha}_{a},\Psi^{\prime},\overline{\Psi}^{\prime})
 \label{VI9}
 \end{equation}
Notice that after the conformal 
transformation (\ref{VI4})-(\ref{VI6}), the $\sigma$-contribution to the 
spin-connection (the second term in the r.h.s. of Eq. (\ref{C4})) is 
canceled. 

Notice also that once again we performed the rescaling (\ref{V15}) in 
order to provide the standard form of the appropriate equations for the 
gauge and gravitational fields.

\bigskip
\subsection{Nambu - Jona-Lasinio model in the NGVE theory}

\bigskip

As it is shown in Appendix B, the famous Nambu - Jona-Lasinio (NJL) model 
\cite{NJL} works in a special way in the context of the NGVE theory: in 
the theory with only fermionic matter, the constraint does not impose 
restrictions on dynamics (see also \cite{GK2}).

If we generalize the model towards a realistic theory by the 
consideration both bosonic and fermionic sectors as in the previous 
subsection, then still the fermionic part does not contribute to the 
constraint if we restrict ourself to the NJL - type model.
In this case the constraint works exactly as in Sec. V where bosonic 
sector breaks the LES explicitly and the constraint becomes an equation 
for $\chi$.

In the Einstein frame, the NJL term (see the last term 
in Eq. (\ref{VI7})) remains its original form in terms of the transformed 
fields ${\Psi}^{\prime}$ and $\overline{\Psi}^{\prime}$. 

As it is well known \cite{Nambu-Bardeen}, the the NJL mechanism 
allows for a dynamical mass generation in realistic models. Since in the 
Einstein frame the 
theory takes the canonical form, therefore the same mechanisms of the 
dynamical mass generation can be applied also here.

\bigskip
\subsection{Classical model for generating mass of fermions}

\bigskip 

As we have discussed in the previous subsection, there is a possibility 
to generate masses of fermions in the NGVE theory due to the quantum 
effect in the NJL model. Now we are going to show that the NGVE theory in 
the context of the model with the action (\ref{VI1}) allows for a purely 
classical mechanism of obtaining fermion masses.

The constraint (\ref{B7}) corresponding to the model (\ref{VI1}) is
\begin{equation}
\overline{\Psi}\Psi U^{\prime}(\overline{\Psi}\Psi)-
2U(\overline{\Psi}\Psi)-2\left [m^{4}(2uf^{\prime}(u)-f(u))-
(V(\varphi)+M)\right ]=0
 \label{VI10}
 \end{equation}   
where Eq. (\ref{B10}) has been used.

Directing our attention to the situation where the scalar field 
excitations around the vacuum \{$u=u_{0}$, $\varphi=\varphi_{0}$ with 
condition (\ref{V27})\}, (see Sec.VB) are ignored, we get the 
constraint for the first order fluctuations of $\Psi$ and $\overline{u}
\equiv u-u_{0}$
\begin{equation}
\overline{\Psi}\Psi U^{\prime}(\overline{\Psi}\Psi)-
2U(\overline{\Psi}\Psi)-
4m^{4}u_{0}f^{\prime\prime}(u_{0})\,\overline{u}=0
 \label{VI11}
 \end{equation}
where Eq. (\ref{V26}) has been used.

We consider now the  model with  fermionic selfinteraction of the form
\begin{equation}
U(\overline{\Psi}\Psi)=-C(\overline{\Psi}\Psi)^{q}, \qquad C=km^{4-3q}
 \label{VI12}
 \end{equation}
where $k>0$ is a dimensionless parameter. 

Remembering that our aim is to study the theory in the Einstein frame, we 
first calculate $\chi$. From Eq. (\ref{V21}) we have $\chi=
\frac{\omega}{f^{\prime}}$ and expanding $f^{\prime}$ around $u=u_{0}$ we get
$f^{\prime}=f^{\prime}(u_{0})+f^{\prime\prime}(u_{0})\overline{u}+\ldots =
f^{\prime\prime}(u_{0})\overline{u}$. $\overline{u}$ can be obtained from 
Eq. (\ref{VI11}) and for special form (\ref{VI12}) we obtain, after a 
simple computation, the following equation for $\chi$
\begin{equation}
\frac{\omega}{\chi}=\frac{C}{4u_{0}m^{4}}\left 
[(2-q)\chi^{q/2}(\overline{\Psi}^{\prime}\Psi^{\prime})^{q}\right ]
 \label{VI13}
 \end{equation}
in terms of the fermion field $\Psi^{\prime}$ in the Einstein frame (see 
Eq. (\ref{VI5})) that is
\begin{equation}
\chi=\left (\frac{C(2-q)}{4\omega u_{0}m^{4}}\right )^{-\frac{2}{q+2}}
\left (\overline{\Psi}^{\prime}\Psi^{\prime}\right )^{-\frac{2q}{q+2}}
 \label{VI14}
 \end{equation}
 
Finally, we observe that in the Einstein frame, the last term in Eq. 
(\ref{VI7}) has the form of a mass term, with a mass $m_{f}$ given by
\begin{equation}
m_{f}=Cq\chi^{-1/2}
 (\overline{\Psi}\Psi)^{q-1}=Cq\chi^{\frac{q-2}{2}}
(\overline{\Psi}^{\prime}\Psi^{\prime})^{q-1}
 \label{VI15}
 \end{equation}

Such a term will be a legitimate mass term only if $m_{f}$ as given by 
Eq. (\ref{VI15}) is a genuine constant. From Eqs. (\ref{VI14}) and 
(\ref{VI15}) we obtain that $m_{f}$ is given by 
\begin{equation}
m_{f}=Cq\left (\frac{C(2-q)}{4\omega u_{0}m^{4}}\right )^{\frac{2-q}{q+2}}
(\overline{\Psi}^{\prime}\Psi^{\prime})^{\frac{3q-2}{q+2}}
 \label{VI16}
 \end{equation}

We observe therefore that $m_{f}$ is indeed a genuine constant if 
$q=2/3$:
\begin{equation}
m_{f}=\frac{mk^{3/2}}{\sqrt{3\omega u_{0}}}
 \label{VI17}
 \end{equation}

We conclude therefore that if we start with a  fermion selfinteraction 
term in the original Lagrangian density (see Eq. (\ref{VI1})) of the form
\begin{equation}
U(\overline{\Psi}\Psi)=-C(\overline{\Psi}\Psi)^{2/3},
\label{VI18}
 \end{equation}
 we obtain normal 
propagation of a massive fermion in the (physical) Einstein frame.

One should notice at this point that a selfinteraction of the form 
(\ref{VI18})
 has a remarkable feature: such a term 
comes in the action with a coupling constant $C=km^{2}$ with 
dimensionality $mass^{2}$, just as we are used in the case of bosonic 
masses, like in the case of a vector meson mass term for example.

Notice that the same result, that is the fact that the form of the 
fermion selfinteraction (\ref{VI18})  provides mass generation of 
fermions in the Einstein frame, is obtained also in the model with 
critical limit (see Sec. VA). In fact, if we start with the Lagrangian 
density similar to Eq. (\ref{V2}) and reformulate it in the vierbein - 
spin-connection formalism with adding the fermionic sector as it 
was done in Sec. VIA, then in the limit $p\rightarrow\infty$, the 
selfinteraction (\ref{VI18}) in the Einstein frame turns out to be the 
mass term of fermion where again $m_{f}\propto\frac{m}{\sqrt{\omega}}$.

\bigskip
\section{Unified gauge theories in the context of NGVE theory}

\bigskip 

Now we able to formulate a realistic gauge theory. For the illustration 
of this we take $SU(2)\times U(1)$ model of electroweak interaction. 
However, there are no obstacles to formulate QCD, GUT or any other 
spontaneously broken gauge model in the context of the 
NGVE theory. The common feature of such models in the NGVE theory is that 
{\em the spontaneous symmetry breaking (SSB) in the vacuum 
$\{\varphi_{0},y_{0}\}$  does not 
generate the vacuum energy and therefore the cosmological constant is 
equal to zero in this vacuum}.

The $SU(2)\times U(1)$ model of electroweak interaction in the NGVE 
theory has to be considered together with  gravitational interaction and 
the Lagrangian density is the following
\begin{eqnarray}
L=-\frac{1}{\kappa}V^{a\mu}V^{b\nu}R_{\mu\nu 
ab}(\omega)-m^{4}f(u)+
i\,\overline{L}\not\!\!DL+i\,\overline{e}_{R}\not\!\!De_{R}+
i\,\overline{\nu}_{R}\not\!\!D\nu_{R}
\nonumber\\
+|D_{\mu}\varphi|^{2}-V(|\varphi|)
-\lambda_{e}m^{4/3}(\overline{L}\,e_{R}\,\varphi+h.c.)^{2/3}
\label{VII18}
 \end{eqnarray}

In Eq. (\ref{VII18}) we used notations of Ref. \cite{Oku}:
$SU(2)$ vector gauge field 
$\tilde{\vec{A}}_{\mu}=(\tilde{A}_{1\mu},\tilde{A}_{2\mu},\tilde{A}_{3\mu})$;
\quad $U(1)$ abelian gauge field $\tilde{B}_{\mu}$;\quad Left lepton 
dublet \begin{displaymath}
L=\left( \begin{array}{c}
\nu_{L}\\
e_{L}
\end{array} \right);
\end{displaymath}
Right $SU(2)$ singlets $\nu_{R}$ and $e_{R}$;
$SU(2)$ scalar fields dublet
\begin{displaymath}
\varphi=\left( \begin{array}{c}
\varphi^{+}\\
\varphi^{0}
\end{array} \right)
\end{displaymath}
and corresponding antiparticles $\varphi^{\dag}=(\varphi^{-}, 
\tilde{\varphi^{0}})$. The 
left and right components of 
fermions are defined by $\Psi_{L}\equiv\frac{1}{2}(1+\gamma_{5})\Psi$ and
$\Psi_{L}\equiv\frac{1}{2}(1-\gamma_{5})\Psi$ correspondingly;
$D_{\mu}=\partial_{\mu}-ig \vec{T} \tilde{\vec{A}}_{\mu}-
ig^{\prime}\frac{Y}{2}\tilde{B}_{\mu}$ where the hypercharge $Y$ is:
$Y=-1$ for $\nu_{L}$ and $e_{L}$; \quad  $Y=-2$ for $e_{R}$.\quad  
$\vec{T}=\frac{1}{2}\vec{\tau}$\quad  for \quad $\varphi$ and $L$. For 
isoscalar fields \, $e_{R}$ \quad and\quad $\nu_{R}$,\quad $T=0$. 
The last term in (\ref{VII18}) is written in the form which provides us 
the mechanism for the fermion mass generation described in Sec. VIC and 
accompanied by SSB. Parameter $\lambda_{e}$ in the last term of Eq. 
(\ref{VII18}) is
dimensionless coupling constant.

Operator $\not\!\!D$ in the third term $\overline{L}\not\!\!DL$ of Eq. 
(\ref{VII18}) is defined as follows:
\begin{equation}
\not\!\!D\equiv\overrightarrow{\not\!\!D}_{L}-\overleftarrow{\not\!\!D}_{L}
\label{VII19}
 \end{equation}
where
\begin{equation}
\overrightarrow{\not\!\!D}_{L}\equiv
\frac{1}{2}V^{\mu}_{a}\gamma^{a}\left(\vec{\partial}_{\mu}
+\frac{1}{2}\omega_{\mu}^{cd}I\sigma_{cd}-
\frac{i}{2}\tilde{g}\vec{\tau}\tilde{\vec{A}}_{\mu}
+\frac{i}{2}\tilde{g}^{\prime}\tilde{B}_{\mu}\right)
\label{VII20}
 \end{equation}
\begin{equation}
\overleftarrow{\not\!\!D}_{L}\equiv\frac{1}{2}
\left(\overleftarrow{\partial}_{\mu}
-\frac{1}{2}\omega_{\mu}^{cd}I\sigma_{cd}+
\frac{i}{2}\,\tilde{g}\vec{\tau}\tilde{\vec{A}}_{\mu}
-\frac{i}{2}\tilde{g}^{\prime}\tilde{B}_{\mu}\right)\gamma^{a}V_{a}^{\mu}
\label{VII21}
 \end{equation}
where $I$ is $2\times 2$ unit matrix in the isospin space.
The forth and the fifth terms in Eq. (\ref{VII18}) are defined by equations:
\begin{equation}
\overline{e}_{R}\not\!\!De_{R}\equiv\overline{e}_{R}
(\overrightarrow{\not\!\!D}_{eR}-
\overleftarrow{\not\!\!D}_{eR})e_{R}
\label{VII22}
 \end{equation}
\begin{equation}
\overrightarrow{\not\!\!D}_{eR}\equiv\frac{1}{2}V^{\mu}_{a}\gamma^{a}
\left(\vec{\partial}_{\mu}
+\frac{1}{2}\omega_{\mu}^{cd}\sigma_{cd}
+i\tilde{g}^{\prime}\tilde{B}_{\mu}\right)
\label{VII23}
 \end{equation}
\begin{equation}
\overleftarrow{\not\!\!D}_{eR}\equiv\frac{1}{2}
\left(\overleftarrow{\partial}_{\mu}
-\frac{1}{2}\omega_{\mu}^{cd}\sigma_{cd}+
-i\tilde{g}^{\prime}\tilde{B}_{\mu}\right)\gamma^{a}V_{a}^{\mu}
\label{VII24}
 \end{equation}
\begin{equation}
\overline{\nu}_{R}\not\!\!D\nu_{R}\equiv
\overline{\nu}_{R}(\overrightarrow{\not\!\!D}_{\nu R}-
\overleftarrow{\not\!\!D}_{\nu R})\nu_{R}
\label{VII25}
 \end{equation}
\begin{equation}
\overrightarrow{\not\!\!D}_{\nu R}\equiv\frac{1}{2}V^{\mu}_{a}\gamma^{a}
\left(\vec{\partial}_{\mu}
+\frac{1}{2}\omega_{\mu}^{cd}\sigma_{cd}\right)
\label{VII26}
 \end{equation}
\begin{equation}
\overleftarrow{\not\!\!D}_{\nu R}\equiv\frac{1}{2}
\left(\overleftarrow{\partial}_{\mu}
-\frac{1}{2}\omega_{\mu}^{cd}\sigma_{cd}\right)\gamma^{a}V_{a}^{\mu}
\label{VII27}
 \end{equation}

In the kinetic term of the scalar fields in Eq.(\ref{VII18}) the 
notations are the following:
\begin{equation}
|D_{\mu}\varphi|^{2}\equiv 
(D_{\mu}\varphi)^{\dag}_{i}(D_{\nu}\varphi)^{i}g^{\mu\nu}
 \label{VII28}
 \end{equation}
\begin{equation}
D_{\mu}\varphi\equiv
\left(\partial_{\mu}
-\frac{i}{2}\tilde{g}\vec\tau\tilde{\vec A}_{\mu}
-\frac{i}{2}\tilde{g}^{\prime}\tilde{B}_{\mu}\right)\varphi
\label{VII29}
 \end{equation}

The gauge complex $u$, which enters into Eq. (\ref{VII18}) as an argument
of the function $f$ in the second term, is defined now as follows:
\begin{equation}
u\equiv\frac{1}{4}\tilde{F}_{\mu\nu}\tilde{F}^{\mu\nu}+
\frac{1}{2}Tr\tilde{G}_{\mu\nu}\tilde{G}^{\mu\nu}+
\frac{m^{2}}{\sqrt{-g}}\varepsilon^{\mu\nu\alpha\beta}\partial_{\mu}
A_{\nu\alpha\beta}
\label{VII30}
 \end{equation}
where $\tilde{F}_{\mu\nu}\equiv\partial_{\mu}\tilde{B}_{\nu}-
                               \partial_{\nu}\tilde{B}_{\mu}$, \ 
$\tilde{G}_{\mu\nu}\equiv\partial_{\mu}\tilde{A}_{\nu}-
                               \partial_{\nu}\tilde{A}_{\mu}-
            i\tilde{g}[A_{\mu}A_{\nu}-A_{\nu}A_{\mu}]$ where
$A_{\mu}=\vec{A}_{\mu}\vec{T}$.

From the action (\ref{VII18}) we will come to equations of motion and the 
constraint similar to what we got before but with the appropriate 
modifications related to  the  $SU(2)\times U(1)$ theory. Nonabelian 
structure of the tensor $\tilde{G}_{\mu\nu}$ does not change the 
constraint since nonlinear term of $\tilde{G}_{\mu\nu}$ has the same 
degree of homogeneity in $g^{\mu\nu}$ as the linear terms. After we will 
get Eqs.(\ref{V19}), (\ref{V24}) and (\ref{V25}), we see that 
$\frac{dV_{eff}}{d\varphi}=0$ (without a fine tuning of the original 
potential $V(\varphi)$), may be realized when 
$|\varphi|^{2}=\varphi^{2}_{0}=\frac{1}{2}\eta^{2}$ for some vacuum 
expectation value 
\begin{displaymath}
\left<\varphi\right>=\frac{1}{\sqrt{2}}
\left( \begin{array}{c}
0\\
\eta
\end{array} \right)
\end{displaymath}
\label{VII31}
Then the appropriate gauge complex condensate $u_{0}$ is determined by 
Eq. (\ref{V27}) in such a way that the effective cosmological constant 
defined by Eq. (\ref{V25}) is equal to zero.

{\em Around this vacuum} the constraint and equations of motion take the 
standard form. It is important to note a few remarkable features of the 
theory. First, in spite of the fact that more than one gauge vector field 
enter in the gauge complex $u$ (Eq. (\ref{VII30})), the equation similar to 
Eq. (\ref{V21}) again is enough to provide in the Einstein frame, the form 
similar to Eq. (\ref{V11}) for all of the gauge field equations. Second,   
once again, in order to provide the form (\ref{V11}) of the gauge field 
equations, we have to perform the rescaling
\begin{eqnarray}
\vec A_{\mu}\equiv 2\sqrt{\omega}\tilde{\vec A}_{\mu}; \qquad
B_{\mu}\equiv 2\sqrt{\omega}\tilde{B}_{\mu};\nonumber\\
g=\frac{\tilde{g}}{2\sqrt{\omega}}; \qquad 
g^{\prime}=\frac{\tilde{g}^{\prime}}{2\sqrt{\omega}}
 \label{VII32}
\end{eqnarray}
where the integration constant $\omega$ appears as the universal parameter.
It is interesting that with this rescaling, where 
$\tilde{g}\tilde{\vec A}_{\mu}=g\vec A_{\mu}$, 
$\tilde{g}^{\prime}\tilde{B}_{\mu}=g^{\prime}B_{\mu}$, we obtain 
the canonical form of all of the equations similar to Eqs. (\ref{V10}) - 
(\ref{V13}) with the appropriate modification to the $SU(2)\times U(1)$ 
theory. From equations similar to Eq. (\ref{V11}) we obtain masses of 
vector bosons. After the standard field redefinition from 
$\vec A_{\mu}$, $B_{\mu}$ to intermediate vector bosons 
$W^{+}_{\mu}$, $W^{-}_{\mu}$, $Z_{\mu}$ and electromagnetic field 
$A_{\mu}$, we obtain the following expressions for their masses:
\begin{equation}
m_{W}=\frac{1}{2}g\eta =\frac{\tilde{g}}{4\sqrt{\omega}}\eta; \qquad
m_{Z}=\frac{1}{2}\overline{g}\eta; \qquad  
m_{A}=0,
\label{VII33}
 \end{equation}
where
\begin{equation}
\overline{g}=\sqrt{g^{2}+g^{\prime 2}} = 
\frac{1}{2\sqrt{\omega}}\sqrt{\tilde{g}^{2}+\tilde{g}^{\prime 2}} 
\label{VII34}
 \end{equation}

Turning now to the fermionic sector, once again we can find the 
spin-connection $\omega_{\mu}^{cd}$. Varying the action with respect to 
$\omega_{\mu}^{cd}$ we obtain an equation similar to Eq. (\ref{C2}) and 
the appropriate modification of Eqs. (\ref{C4})-(\ref{C7}) as the 
solution of it. In the Einstein frame the $\sigma$-contribution
$K_{\mu}^{cd}(\sigma)$ disappears as we discussed at the end of Sec. VIA.  
Contribution of the fermions  selfinteraction to $\omega_{\mu}^{\prime ab}$, 
like 
$\propto\kappa\eta_{ci}V_{d\mu}^{\prime}
\varepsilon^{abcd}\overline{e_{L}^{\prime}}
\gamma^{5}\gamma^{i}e_{L}^{\prime}$ 
\quad ($e_{L}^{\prime}$ is the left electron spinor in the Einstein frame)
which are supressed by factor $\kappa=16\pi G$ 
in comparison to the first term in Eq. (\ref{VI9}), may be neglected 
if we are interested in particle physics at energies much less than the 
Planck energy scale. 
Therefore neglecting the last term in Eq. (\ref{VI9}) we remain only 
with the Riemannian contribution to the spin-connection.  

Concerning the mass generation of fermions, we have to point out that the 
choice of the last term in Eq. (\ref{VII18}) is related to our intention 
to use the mechanism for the electron mass generation developed in 
Sec.VIC, where we have found out that the exponent $q$ must be equal to 2/3. 
Comparing 
with notations of Sec.VIC we see that after the SSB, 
the factor $C$ in Eq. (\ref{VI18}) has to be identified in the low energy 
theory as 
\begin{equation}
km^{2}=C=\lambda_{e}\left(\frac{\eta}{\sqrt{2}}\right)^{2/3}m^{4/3}
\label{VII35}
 \end{equation}
As a result we get from Eq. (\ref{VI17}):
\begin{equation}
m_{e}=\frac{\lambda_{e}^{3/2}}{\sqrt{6u_{0}\omega}}\eta
\label{VII36}
 \end{equation}

Notice that ratios between masses of all particles of the model (see Eq. 
(\ref{V28}) for the Higgs boson mass,\, Eqs. (\ref{VII33})), (\ref{VII34}) 
for $W^{\pm}_{\mu}$ and $Z_{\mu}$ masses and the electron mass, Eq. 
(\ref{VII36})) are $\omega$-independent. The same is true for the weak angle
\begin{equation}
\sin\theta_{W}=\frac{g^{\prime}}{\overline{g}}
\label{VII37}
 \end{equation}

It is very interesting that a big value of the integration constant 
$\omega$ pushes gauge coupling constants and all masses to small values. In 
addition, the constant $g_{Y}$ of the effective Yukawa coupling 
$g_{Y}\overline{L}^{\prime}e^{\prime}_{R}\varphi$
\begin{equation}
g_{Y}=\frac{\lambda^{3/2}_{e}}{\sqrt{3u_{0}\omega}}
\label{VII38}
 \end{equation}
has an additional factor $u_{0}^{-1/2}$, which can explain (if the gauge 
complex condensate $u_{0}=\frac{y_{0}}{m^{4}}$ is also big) the observed 
further suppression of the effective Yukawa coupling and therefore the 
appropriate suppression of the observed lepton masses. We can think about 
all these effects related to a big values of $\omega$ and $u_{0}$ as a 
{\em "cosmological seesaw mechanism"}, where masses are driven to small 
values due to the appearance of large number in a denominators.

As we pointed out at the beginning of this section, other gauge unified 
theories can be formulated in the same fashion. If  for example we would  
consider also QCD, then the same effect of additional suppression would 
be  obtained for the masses of quarks.

\bigskip
\section{Discussion and Conclusions}

\bigskip

In this paper we have seen that the idea to allow the measure to be 
determined dynamically rather than postulating it to be $\sqrt{-g}$ from 
the beginning, has deep consequences. In fact, in the context of the 
first order formalism
the NGVE theory does not have a cosmological constant problem.

In this theory, if a four 
index field strength is introduced it develops a condensate which turns 
out to be expressed in terms of other fields. A consequence of this is 
the possibility to produce the standard particle physics dynamics. 
The 
resulting dynamics has then interesting consequences in what concerns to 
the hierarchy problem. As we have seen, all masses and gauge coupling 
constants are driven to small values if the integration constant $\omega$, 
that parametrizes the condensate
strength  (see Eq. (\ref{V21})), is big. In addition to 
this,  masses of fermions are 
driven to small values in comparison with masses of bosons as the gauge 
complex condensate $y_{0}$ becomes big (see Eq. (\ref{VII38})).

The appearance of the parameter $\omega$ in the relation between the 
original coupling constants and the effective ones 
suggests an idea that it may be possible to think in different ways 
concerning renormalization theory. It seems to promise allure prospects 
since the strength of the condensate  specifies a boundary condition or 
state of the Universe.

It is very interesting that a theory designed in order to solve the 
cosmological constant problem tell us about an apparently unrelated subject, 
like what determines the effective coupling constants and masses of the 
theory. One should recall that the wormhole approach to the 
cosmological constant problem ends up claiming that wormholes determine 
all couplings of the theory also \cite{worm}. 

Other important consequence of the theory described in this paper is 
that one can obtain  scalar field dynamics which allows for an 
inflationary era, the possibility of reheating after scalar field 
oscillations and the setting down to a zero cosmological constant phase 
at the later stages of cosmological evolution, without fine tuning. It is 
interesting that the theory not just reproduces all possibilities well 
known in the cosmology of the very early universe solving at the same 
time the cosmological constant problem. In addition to this, the effective 
scalar 
field potential includes integration constants - the feature which let 
us to hope that by an appropriate choice of those constants, the correct 
density perturbations and reheating could be obtained naturally. Moreover, 
the integration constant $\omega$ enters both in the effective potential 
and in the effective coupling constants and masses. This means that may 
be  a strong correlation between coupling constants and masses of particles 
and some of the cosmological parameters.

Some open problems are of course apparent. For example, in what appears 
to be the most attractive scheme for generating standard gauge field 
dynamics, scalar potentials and fermion masses, namely the "persistent 
gauge field condensate scenario" (see Sec. VB) has to be understood in a 
deeper way. There a nonspecified function $f$ of a special combination $y$ 
(Eqs. 
(\ref{V1}) or (\ref{VII30}) as examples) of all gauge fields , including 
3-index potential, is 
introduced. For this function we require only the existence of an 
extremum at some point $y=y_{0}>0$ and this is enough to get the
effective action of electro-weak, QCD and other gauge unified models. The 
possible origin of such structure 
as well as the choice of function has to be studied. 

The fact that 
similar type of function appear for example in the QCD effective action as 
result of  radiative corrections \cite{QCDcondens}, is  encouraging. In  
such a case no four index field strength is introduced however and the 
effective action is a function of $F_{a\mu\nu}F_{a}^{\mu\nu}$ there. 
Notice that appearance of a four index field strength condensate in the 
framework of the theory developed in this paper, makes the Lorentz 
invariance of the vacuum in QCD not a problem, as opposed to the 
situation where only regular gauge fields are present as the argument of the 
nontrivial 
function, leading to an expectation value of $F_{a\mu\nu}F_{a}^{\mu\nu}$

As we have seen, the four index field strength plays the central role in 
the model described in this paper. In this connection, one has to recall 
that four index field strength plays a fundamental role in some 
supergravity models, in particular the $D=11$, $N=1$ supergravity and in 
the $D=4$, $N=8$ supergravity theories. The possibility of incorporating 
some versions of supergravity into the framework developed in this paper 
seems therefore a subject which could be a potentially fruitful one.    
 
Finally, another subject that can certainly be studied concerns the 
possibility of exploiting the correlations between the masses and 
coupling constants and the strength of the condensate. If we allow for a 
coupling between the 3-index potential $A_{\mu\nu\alpha}$ and $2+1$ 
dimensional membranes of the form \cite{FF2}
$T\int A_{\mu\nu\alpha}dx^{\mu}\land dx^{\nu}\land dx^{\alpha}$, one can 
see that the condensate strength can suffer a discontinuous change across 
the membrane. This effect can be exploited to construct a "bag" containing 
particles which are very light and weakly 
coupled inside the bag while 
being very heavy and strongly coupled outside the bag. The analogy with 
the famous MIT bag - model \cite{MIT} is self-evident of course.
  
Fynally, an important question that has to be addressed is of course the 
quantization of the theory and whether quantum corrections will respect 
the basic structure of the theory which enables the theory not to have a 
cosmological constant problem. As we have seen, the key of the resolution 
of the cosmological constant problem is based on the study of an action 
of the form (\ref{Action}), where the measure fields $\varphi_{a}$ does 
not enter in the Lagrangian. This form appears then to be associated with 
the existence of the infinite dimensional symmetry (\ref{LP}) which is 
valid if and only if the structure (\ref{Action}) is mantained. We 
interpret this as a strong indication that the resolution of the 
cosmological constant problem discussed in this paper will survive 
quantum corrections, provided no quantum anomalies of the symmetry 
(\ref{LP}) are found. 

 \bigskip
\section{Acknowledgments}
We thank Profs. J. Bekenstein, A. Davidson, F. Hehl and Y. Ne'eman for 
useful and encouraging discussions on the subjects of this paper.

\bigskip

\appendix

\section{Metric - Affine formalism in the NGVE theory and $\lambda$-symmetry}

\bigskip
In Sec. IIIA we have shown that in the NGVE theory there is a 
contribution of the $\chi$-field to the connection (see Eqs. (\ref{GAM2}) 
and (\ref{S2})). This contribution is defined up to the 
$\lambda$-symmetry transformation (\ref{Gamal}). By using this symmetry, 
in Sec.IIIA we have chosen the gauge where the antisymmetric part of the 
connection (that is a $\chi$-contribution into the torsion) disappears.

It is interesting to see what is the geometrical meaning of the 
$\lambda$-gauge dependent contribution to the connection (\ref{GAM2}), 
(\ref{S2}). For this we calculate the covariant derivative of the 
metric tensor $g_{\mu\nu}$ with the connection defined by (\ref{GAM2})
and (\ref{S2}) and we get
\begin{equation}
g_{\mu\nu ;\alpha}=-2g_{\mu\nu}\lambda_{,\alpha}\equiv N_{\mu\nu\alpha}.
        \label{A1}
\end{equation}
This means that the $\lambda$ dependent contribution to the connection
is responsible for the appearance of the nonmetricity tensor \cite{nonmetr}.

With the choice $\lambda = \frac{1}{2}\sigma$, in Sec. IIIA, we have 
eliminated the $\chi$-contribution into the torsion keeping the 
nonmetricity tensor $N_{\mu\nu\alpha}$ which in this 
"$\sigma$-torsionless" gauge is equal to \quad 
$-g_{\mu\nu}\sigma_{,\alpha}$. However we have the freedom to choose for 
example the "$\sigma$- metric - compatible" gauge where the nonmetricity 
disappears: $\lambda =constant$. In such a case, the torsion is not 
eliminated from the connection (\ref{GAM2}), (\ref{S2}).

 Notice that these peculiarities of the $\lambda$-symmetry concerning the 
possibility of eliminating the $\chi$-contribution to the torsion or, 
alternatively, of eliminating the $\chi$-contribution to the nonmetricity 
appear to be a very interesting feature of the NGVE theory in the metric 
- affine formalism. This feature results from the basic assumption that 
not only metric and connection are independent dynamical variables (as it 
is in the case of the Metric - Affine theory), but also the measure degrees 
of freedom are independent variables when varying the action. This is why 
we will call this theory the Metric - Affine - Measure (MAM) theory.

\bigskip
\section{Constraint and LES in the Vierbein - Spin-Connection  
formalism}

\bigskip

To incorporate fermions into the NGVE theory we have to use the vierbein 
- spin-connection (VSC) formalism (see, however a purely affine approach 
due to Ne'eman \cite{Neeman}). In this Appendix we review in a short form 
how the NGVE principle works in the VSC-formalism.

 In this case we define \cite{Gasp} 
\begin{equation} 
R(\omega ,V) =V^{a\mu}V^{b\nu}R_{\mu\nu ab}(\omega),
\label{B1}
\end {equation}
\begin{equation}
R_{\mu\nu ab}(\omega)=\partial_{\mu}\omega_{\nu ab}
-\partial_{\nu}\omega_{\mu ab}+(\omega_{\mu a}^{c}\omega_{\nu cb}
-\omega_{\nu a}^{c}\omega_{\mu cb})
        \label{B2}
\end{equation}
where $V^{a\mu}=\eta^{ab}V_{b}^{\mu}$, $\eta^{ab}$ is the diagonal
$4\times 4$
matrix with elements $(1, -1,-1,-1)$ on the diagonal, $V_{a}^{\mu}$
are the vierbeins and $\omega_{\mu}^{ab}=-\omega_{\mu}^{ba}$ 
($a,b=0,1,2,3$) is
the spin connection. The matter Lagrangian $L_{m}$ that appears in
Eq.(\ref{Action}) is now a function of matter fields, vierbeins and spin
connection, considered as independent fields. We assume for simplicity
that $L_{m}$ does not depend on the derivatives of vierbeins and spin
connection.

We are now going to study the theory defined by the action 
(\ref{Action}) in the case that the scalar curvature is defined by 
(\ref{B1}),(\ref{B2}).

As in Sec. II, variation with respect to the scalar fields $\varphi_{a}$ 
leads to the equations

\begin{equation}
 A_{a}^{\mu}\partial_{\mu}(-\frac{1}{\kappa}R(V,\omega ) + L_{m}(V,\omega 
, matter fields)) = 0
 \label{B3}
 \end{equation}
which implies, if $\Phi \neq 0$, that

\begin{equation}
 -\frac{1}{\kappa}R(V,\omega ) + L_{m}(V,\omega, matter fields) = M
 \label{B4}
 \end{equation}

On the other hand, considering the equations obtained from the variation 
of the vierbeins, we get if  $\Phi \neq 0$
\begin{equation}
 -\frac{2}{\kappa}R_{\mu a}(V,\omega)
+\frac{\partial L_{m}}{\partial V^{a\mu}} = 0,
 \label{B5}
 \end{equation}
where
\begin{equation}
 R_{\mu a}(V,\omega)\equiv 
V^{b\nu}R_{\mu\nu ab}(\omega).
\label{B6}
\end{equation}

Notice that eq.(\ref{B5}) is indeed invariant under the shift 
$L_{m}\rightarrow L_{m}+const$. By using Eq. (\ref{B1}) we can eliminate 
$R(\omega)$ from the equations (\ref{B4}) and (\ref{B5}) after contracting 
the last one with $V_{a\mu}$. As a result we obtain {\em the nontrivial 
constraint} in the form 

\begin{equation}
V^{a\mu}\frac{\partial(L_{m}-M)}{\partial V^{a\mu}}-2(L_{m}-M)=0
\label{B7}
 \end{equation}
which replaces Eq. (\ref{II3}) (see Sec. II) and in the absence of 
fermions, is equivalent to the constraint (\ref{II3}).

The simplest example of a fermion is that of spin 1/2 particles. In
this case we regard the spinor field $\Psi$  as a general
coordinate
scalar and transforming nontrivially with respect to local Lorentz
transformation according to the spin $1/2$ representation of the Lorentz
group.

The NGVE principle prescripts for the fermionic  hermitian action (which 
allows for the possibility of fermion self interactions) to be of the form
\begin{equation}
S_{f}=\int L_{f}\Phi d^{4}x
 \label{B8}
 \end{equation}
where
\begin{equation}
 L_{f}=\frac{i}{2}\overline{\Psi}
[\gamma^{a}V_{a}^{\mu}(\vec\partial
_{\mu}+\frac{1}{2}\omega_{\mu}^{cd}\sigma_{cd})
-(\overleftarrow{\partial}_{\mu}-\frac{1}{2}\omega_{\mu}^{cd}\sigma_{cd})
\gamma^{a}V_{a}^{\mu}]\Psi+U(\overline{\Psi}\Psi)
 \label{B9}
 \end{equation}
Here $\sigma_{cd}\equiv \frac{1}{4}[\gamma_{c},\gamma_{d}]$.
Spin-connection $\omega_{\mu}^{cd}$ should be determined by the equation 
obtained
from the variation of the full action with respect to $\omega_{\mu}^{cd}$ 
(see Appendix C). 

From (\ref{B9}) and using the equations of motion 
derived from the action (\ref{B8}),(\ref{B9}), we get
\begin{equation}
V_{a}^{\mu}\frac{\partial L_{f}}{\partial V_{a}^{\mu}}-2L_{f}=
\overline{\Psi}\Psi U'-2U,
 \label{B10}
 \end{equation}
where $U'$ is the derivative of $U$ with respect to its argument 
$\overline{\Psi}\Psi$. We see
that the constraint (\ref{B7}) is satisfied on the mass shell (since the
fermion equations of motion are used) with $M=0$ for $L_{f}$
defined by eq.(\ref{B9}) if, for example, $U=c(\overline{\Psi}\Psi)^{2}$.
Any other quartic interaction, like
$\overline{\Psi}\gamma_{a}\Psi\overline{\Psi}\gamma^{a}\Psi$,
$\overline{\Psi}\sigma_{ab}\Psi\overline{\Psi}\sigma^{ab}\Psi$,
$(\overline{\Psi}\gamma_{5}\Psi)^{2}$, etc. would also satisfy the
constraint (\ref{B7}) on the mass shell with $M=0$. In particular, the
Nambu - Jona-Lasinio
model\cite{NJL} would also satisfy the constraint (\ref{B7}) on the mass
shell with $M=0$.

In the presence of Dirac fermions  with the Lagrangian density (\ref{B9}),
the LES  (\ref{ES1}), (\ref{ES2}) is appropriately
generalized to
\begin{equation}
V^{a}_{\mu}(x)=J^{-1/2}(x)V^{\prime a}_{\mu}(x);\qquad V_{a}^{\mu}(x)=
J^{1/2}(x)V^{\prime\mu}_{a}(x)
\label{B11}
\end{equation}
\begin{equation}
\Phi(x)=J^{-1}(x)\Phi^{\prime}(x)
\label{B12}
\end{equation}
\begin{equation}
\psi(x)=J^{1/4}(x)\psi^{\prime}(x);\qquad
\overline{\psi}(x)=J^{1/4}(x)\overline{\psi}^{\prime}(x)
\label{B13}
\end{equation}
provided that $V(\overline{\psi}\psi)$ has a form of one of the mentioned
above quartic interactions.
Notice that in this case the condition for the invariance of the action with
the matter Lagrangian
(\ref{B9}) under the transformations (\ref{B11})-(\ref{B13}) is not just
the simple homogeneity of degree 1 in $g^{\mu\nu}$ or degree 2 in
$V_{a}^{\mu}$, because of the presence of the fermion transformation
(\ref{B13}).

\bigskip
\section{Connection in the VSC-formalism}
\bigskip

We analyze here what is the dependence of the spin connection 
$\omega_{\mu}^{ab}$ on $V^{a}_{\mu}$, $\chi$, $\Psi$ and $\overline{\Psi}$.
Varying the action (\ref{B8}), (\ref{B9}) with respect to 
$\omega_{\mu}^{ab}$ and making use that
\begin{equation}
R(V,\omega)\equiv 
-\frac{1}{4\sqrt{-g}}\varepsilon^{\mu\nu\alpha\beta}\varepsilon_{abcd}
V^{c}_{\alpha}V^{d}_{\beta}R_{\mu\nu}^{ab}(\omega)
\label{C1}
 \end{equation}
we obtain
\begin{equation}
\varepsilon^{\mu\nu\alpha\beta}\varepsilon_{abcd}[\chi 
V^{c}_{\alpha}D_{\nu}V^{d}_{\beta}
+\frac{1}{2}V^{c}_{\alpha}v^{d}_{\beta}\chi,_{\nu}]+
\frac{\kappa}{4}\sqrt{-g}V^{c\mu}\varepsilon_{abcd}\overline{\Psi}
\gamma^{5}\gamma^{d}\Psi=0, 
\label{C2}
 \end{equation}  
where 
\begin{equation}
D_{\nu}V_{a\beta}\equiv\partial_{\nu}V_{a\beta}
+\omega_{\nu a}^{d}V_{d\beta}
\label{C3}
 \end{equation}
The solution of Eq. (\ref{C2}) is represented in the form
\begin{equation}
\omega_{\mu}^{ab}=\omega_{\mu}^{ab}(V) + K_{\mu}^{ab}(\sigma)  + 
K_{\mu}^{ab}(V,\overline{\Psi},\Psi)
\label{C4}
 \end{equation}
where 
\begin{equation}
\omega_{\mu}^{ab}(V)=V_{\alpha}^{a}V^{b\nu}\{ ^{\alpha}_{\mu\nu}\}-
V^{b\nu}\partial_{\mu}V_{\nu}^{a}
\label{C5}
 \end{equation}
is the Riemannian part of the connection,
\begin{equation}
K_{\mu}^{ab}(\sigma)=\frac{1}{2}\sigma_{,\alpha}(V_{\mu}^{a}V^{b\alpha}-
V_{\mu}^{b}V^{a\alpha})
\label{C6}
 \end{equation}
and
\begin{equation}
K_{\mu}^{ab}(V,\overline{\Psi},\Psi)=
\frac{\kappa}{8}\eta_{ci}V_{d\mu}\varepsilon^{abcd}\overline{\Psi}
\gamma^{5}\gamma^{i}\Psi.
\label{C7}
 \end{equation}
Notice that the spin-connection $\omega_{\mu}^{ab}$ defined by Eqs. 
(\ref{C4})-(\ref{C7}) is invariant under the LES transformations 
(\ref{B11})-(\ref{B13}).

\end{document}